\documentclass[11pt]{article}
\usepackage{amssymb}

\begin{document}
\def\be{\begin{equation}}
\def\ee{\end{equation}}
\newcommand{\ds}{\displaystyle}
\newcommand{\Real}{\mathbb{R}}
\newcommand{\Natural}{\mathbb{N}}
\newcommand{\Integer}{\mathbb{Z}}
\newcommand{\Complex}{\mathbb{C}}

\begin{titlepage}

\begin{center}
{\Large Group Theoretical Quantization  of a Phase Space $S^1
\times \Real^+$
\\ and the Mass Spectrum of Schwarzschild Black Holes \vspace{0.2cm}
 \\ in $D$ Space-Time
Dimensions}\\ \vspace{0.7cm} M.\ Bojowald\footnote{E-mail:
bojowald@physik.rwth-aachen.de}, H.A.\ Kastrup\footnote{E-mail:
kastrup@physik.rwth-aachen.de}, F.\ Schramm and T.\
Strobl\footnote{E-mail: pth@tpi.uni-jena.de}\vspace{0.2cm}
\\ {\small Institute for Theoretical Physics, RWTH
Aachen, D-52056 Aachen, Germany } \end{center}\vspace{0.4cm}

 \begin{center}{\bf Abstract}\end{center}  {\small The symplectic
  reduction of pure
spherically symmetric (Schwarz\-schild) classical gravity in $D$
space-time dimensions yields a 2-dimensional phase space of
obser\-vables consisting of the mass $M\; (>0)$ and a canonically
conjugate (Killing) time variable $T$. Imposing (mass-dependent)
periodic boundary conditions in time on the associated quantum
mechanical plane waves which represent the Schwarzschild system
in the period just before or during the formation of a black hole,
yields an energy spectrum of the  hole which realizes the old
Bekenstein postulate that the quanta of the horizon $A_{D-2}$ are
multiples of a basic area quantum.

 In the present paper it is shown that the
 phase space  of such Schwarzschild black holes in $D$ space-time
  dimensions
   is  symplectomorphic  to a symplectic ma\-ni\-fold $\cal S$=$\{
    (\varphi \in
   \Real \bmod 2\pi,~ p~ \propto A_{D-2} \in \Real^+)\}
  $ with the symplectic form $d\varphi \wedge dp$. As the action of
   the group $ SO^{\uparrow}(1,2)$
   on that  manifold is
  transitive,  effective and Hamiltonian,
   it can be used for a group theoretical
   quantization of the system. The area operator $\hat{p}$ for the horizon
   corresponds to the generator of the compact subgroup $SO(2)$
    and becomes quantized accordingly:

    The positive discrete series of the irreducible unitary
   representations of the group $SO^{\uparrow}(1,2)$  yields an (horizon)
   area spectrum $\propto (k+n)$, where $k=1,2,\ldots$ characterizes
   the representation and $n=0,1,2, \ldots $ the number of area quanta.
   If one employs the unitary representations of the
    universal covering group of $SO^{\uparrow}(1,2)$, the number $k$
    can take any fixed positive real value ($\theta$-parameter!).

    The unitary representations of the positive discrete series provide
    concrete Hilbert spaces for quantum Schwarzschild black
    holes.} \\ \\
    PACS numbers: 03.65.Ca; 03.65.Fd; 04.60.Ds; 04.60.Kz
\end{titlepage}
\setcounter{footnote}{0}
\section{Introduction}
Understanding the quantum properties of black holes and the
associated quantum statistics (thermodynamics) is one of the
outstanding topics of present-day approaches to quantum gravity:
In string theory extremal black holes with their relations between
masses and charges and the associated degeneracies play a
prominent role (for a review see Ref. \cite{pe}). In loop quantum
gravity the action of the corresponding area operators on
appropriate spin-network states is expected to yield information
about the entropy of black holes (see the reviews \cite{as1}).

Already in 1974 Bekenstein proposed \cite{be1} a Bohr-Sommerfeld
type of quantization for black holes which amounts to the simple
quantum relation \be A(n) \propto n\; l^2_{P,4}~,~~ n \in \Natural
\equiv \{1,2,\ldots \}~~, \ee ($l_{P,4}$: Planck's length in $D=4$
space-time dimensions) for the 2-dimensional area $A=4\pi\,R_S^2$
 of the horizon of a Schwarzschild black hole
(SBH) in 4 space-time dimensions. Because of $R_S=2\,G\,M/c^2$
this is equivalent to the assertion that the energy levels $E_n$
of such holes are proportional to $\sqrt{n} ~$: \be E_n \propto
\sqrt{n}\; E_{P,4}~,~~ n \in \Natural~~. \ee ($E_{P,4}$: Planck's
energy).

 Together with the assumption that the $n$-th level (1)
has a degeneracy $d_n=g^n,g>1,$ one then gets the
Bekenstein-Hawking entropy of
 the black hole
as proportional to the area of the horizon. As to the further
history of the spectrum (1) and its degeneracies see the recent
review by Bekenstein \cite{be2} and the Refs. \cite{ka2,ka5}.

However, having an appropriate spectrum including its degeneracies
is not sufficient for a complete quantum mechanical description of
the system. For that purpose one has to know the Hilbert space and
the action of the basic operators associated with the system.

It is the main aim of the present paper to provide that Hilbert
space (or a number of unitarily equivalent ones) and the basic
self-adjoint operators in terms of the positive discrete series of
the irreducible unitary representations of the proper
orthochronous Lorentz group $SO^{\uparrow}(1,2)$ or its respective
covering groups.

The method to achieve this makes use of a proposal of one of us
\cite{ka1} how to relate the spectrum (2) to plane wave solutions
of the simple Schroedinger equation resulting
 from the symplectic reduction of spherically symmetric pure
gravity \cite{tk,ku,kun,st}.

In order to have this paper to some extent self-contained we first
summarize the essential steps and assumptions leading to the
required quantum theory. As to more details and possible questions
concerning these steps we refer to Ref.\ \cite{ka1} and the
companion paper \cite{ka5}.

The symplectic reduction of spherically symmetric pure Einstein
gravity yields one pair of canonical variables (``observables'' in
the sense of Dirac), namely the (ADM) mass $M$ of the system ---
here considered as the canonical momentum --- and  a canonically
conjugate time variable $T$, with the associated symplectic form
\be \omega = dT \wedge dM~~. \ee The Schroedinger equation of
 the corresponding
quantum mechanics is extremely simple: \be i \hbar \partial_{\tau}
\psi(\tau) = Mc^2\psi(\tau)~~, \ee where $\tau$ is the proper time
of an observer at (Minkowski flat) spatial infinity.

 Eq. (4) has
the plane wave solutions \be e^{\displaystyle -(i/\hbar)Mc^2\,
\tau}~~. \ee  Up to now no restrictions have been imposed on the
spectrum of the masses $M$ which a priori may be any real number $
M \in \Real$. However, for physical reasons --- no naked
singularities etc. --- one wants $M$ to be a positive real number,
$M \in \Real^+ \equiv \{r \in \Real, r>0 \}$.

 The discrete spectrum (2) may be obtained as
follows \cite{ka1}: Suppose the plane wave (5) represents the
system only during a finite time period $\Delta$. Implementing
this finite duration by periodic boundary conditions in $\tau$
leads to the relation \be c^2 M\; \Delta = 2 \pi \hbar\, n,~n\in
\Natural ~~. \ee Here the assumption $M>0$ is made.

It appears necessary to stress the following point in connection
with the boundary condition (6): The postulate that the wave
function (5) has the period $\Delta$ does not mean that the
asymptotic time $\tau$ is periodic. It just means that the system
is in a (quasi-) stationary state (5) during a finite time
interval $\Delta$. This is a situation completely analogous to a
system of free particles in a finite spatial interval of length
$L$ where the associated state is a plane wave with periodic
spatial boundary conditions (yielding discrete momenta). Such a
property of the wave function does not mean that space itself is
confined to an interval or periodic.

 The
question now is how to choose the time interval $\Delta$.
 As the only intrinsic
length (time) scale of the system is the Schwarzschild radius
$R_S(M)$, the interval $\Delta$ has to be related to $R_S/c$.
There are two important time scales associated with a
Schwarzschild black hole, namely the ``formation'' time or the
inverse Hawking temperature which are proportional to $R_S/c$ and
the ``evaporation'' time due to Hawking radiation which in 4
space-time dimensions is proportional to $R_S^3$
(Stefan-Boltzmann's law \cite{wa1}). It appears to be more
plausible \cite{ka5} to associate the plane wave (5) with the
quasi-stationary pre-collapse phase of the system than with the
evaporation one. The assumption $\Delta = \gamma\, R_S/c\, ,~
\gamma = O(1)\, ,$ leads to the quantization condition \be \gamma
c M_n\, R_S(M_n)= 2 \pi \hbar\,n,~ n\in \Natural~~. \ee With $R_S
=2\,M\,G/c^2$ one immediately get the relations (1) and (2).

 The
relations (5)-(7) may be generalized \cite{ka4,ka5} to arbitrary
space-time dimensions $D\geq 4$ (see also appendix B of the
present paper),
 where the relationship between Schwarzschild mass and
Schwarzschild radius is given by \cite{per1}  \be R_S^{D-3}=
\frac{16 \pi G_D M}{c^2 (D-2)\omega_{D-2} }~~. \ee ($G_D$ denotes
the gravitational constant in $D$ space-time dimensions and
$\omega_{D-2}= 2\pi^{(D-1)/2}/\Gamma((D-1)/2)$ the volume of
$S^{D-2}$.)

 As the area $A_{D-2}$ of the $(D-2)$-dimensional
horizon is  $R_S^{D-2}\omega_{D-2}$, the relations (7) and (8)
imply a horizon area spectrum \begin{eqnarray} A_{D-2}(n) &=& n\,
\tilde{a}_{D-2}~,~n \in \Natural~, \\ &&
 \tilde{a}_{D-2}=\frac{32
\pi^2 G_D \hbar}{\gamma (D-2)c^3}\equiv \frac{32\, \pi^2}{\gamma
(D-2)}\; l_{P,D}^{D-2}~~, \nonumber \end{eqnarray} which,
 according to Eq.\ (8), leads to the energy spectrum
\begin{eqnarray} E_n& = &\alpha_D\, n^{(D-3)/(D-2)}E_{P,D}~,
 \\ & & \alpha_D =
\left(\frac{(2\pi)^{D-4} \,(D-2)\,
\omega_{D-2}}{8\gamma^{D-3}}\right)^{1/(D-2)},
\nonumber \\ &&
E_{P,D}=(c^{D+1}\,\hbar^{D-3}/G_D)^{1/(D-2)}~~. \nonumber \end{eqnarray}
 The energy $E_n$ may be
interpreted \cite{ka3,ka5} as the surface energy of a ``bubble''
of $n$ area quanta $\tilde{a}_{D-2}$.

 The above arguments, which lead to the
spectra
 (9) and (10),
respectively, are unsatisfactory for the following reasons: The
period $\Delta(M)$ of the time variable $\tau$ (or $T$) is a
function of $M$. This means that the phase space of interest here
is the subspace (``wedge'') of the $(M,\tau)-$plane which is bounded
by the positive $M$-axis --- without the origin --- and the curve
$\tau = \Delta(M) = const.~ M^{1/(D-3)} > 0$, where the points $M$
and $\Delta(M)$ have to be identified. This is an unusual phase
space. In addition it is not obvious what is the Hilbert space
associated with the quantized system.

It is the purpose of the present paper to improve the situation by
employing a group theoretical quantization \cite{ish,ste} based on
the group $SO^{\uparrow}(1,2)$ --- the orthochronous proper Lorentz
group in $1+2$ dimensions --- and its irreducible
 unitary representations:

The transformation   \begin{eqnarray} \varphi &=& \frac{\ds 2 \pi}{\ds \Delta}
\; \tau
= \frac{\ds 2\pi c}{\ds \gamma\, R_S(M)}\; \tau~~, \\ p &=& \beta \,
A_{D-2}(M)~,~ \beta
= \frac{\ds \gamma c\, (D-3)}{\ds 32 \pi^2 G_D}~, \end{eqnarray}
is canonical (symplectic):  \be \omega = d\varphi \wedge d p
 = d\tau \wedge dM~~. \ee
As $  \varphi \in \Real \bmod 2\pi$ now, one sees that the phase
space in question
 is
 diffeomorphic to $S^1 \times \Real^+ \simeq \Real^2 -\{0\}$.
 This phase space may be interpreted as ``half'' of the cotangent bundle
 $T^*S^1= \{(\varphi,p)\}$ with the restriction $p>0$.

  The task is then to quantize this
 classical system appropriately. This will be done in the following
 way:

  In section 2 we summarize, following Isham's
 review \cite{ish}, the essential features of the group theoretical
 approach for quantizing a classical symplectic (phase) manifold. In section
 3 we discuss the symplectic, transitive and effective action of
  the 3-dimensional noncompact group $SO^{\uparrow}(1,2)$
  on \be \mbox{$\cal S$}
  = \{(\varphi,p); \varphi \in
 \Real \bmod 2\pi,~ p >0 \} \ee
  by employing
 the 2-fold covering group $SU(1,1)$.
We show which vector fields  are induced on $\cal S$ by the
generators of three independent one-dimensional subgroups of
$SO^{\uparrow}(1,2)$, how the isomorphism between these vector
fields and the corresponding Hamiltonian ones and their Poisson
algebra looks like and which observables (functions) on $\cal S$
correspond to the three selected Lie algebra elements of
$SO^{\uparrow}(1,2)$. It turns out that the above variable $p$
corresponds to the generator of the compact subgroup of
$SO^{\uparrow}(1,2)$.

 As this group is infinitely connected with
first homotopy group $\pi_1 = \Integer$, its covering groups,
especially the universal one, act only almost effectively on $\cal
S$, because the elements of their discrete abelian center leave
every point of $\cal S$ fixed.

 The phase space $\cal S$ is
diffeomorphic to the complex plane with the origin deleted, and
likewise to
 the coset space $SO^{\uparrow}(1,2)/\mbox{N}$, where $\mbox{N}$ is the
nilpotent subgroup in an Iwasawa decomposition of
$SO^{\uparrow}(1,2)$.

 Quantization, which is discussed in section 4,
 consists in passing to the irreducible unitary representations
of $SO^{\uparrow}(1,2)$ where the  basic observables corresponding
to  three Lie algebra elements mentioned above become self-adjoint
operators. The operator representing the generator of the compact
subgroup, i.e.\ the observable $p$, has a discrete spectrum in all
irreducible unitary representations. However, as we want $p
\propto A_{D-2}$ to be positive, only  the unitary representations
of the ``positive discrete'' series  are suitable for our purpose:

 If we denote the
operator corresponding to $p$ by $\hat{p}$, then this operator has
spectra $\propto k+n, n=0,1,2, \ldots$, where the numbers $k=1,2,
\ldots$ characterize the different unitary representations with
the Casimir operator $k(1-k)$. The main difference between the
unitary representations of $SO^{\uparrow}(1,2)$ and those of its
 universal covering group is that for the latter the number $k$ may be any
 positive real number.

 The irreducible unitary
representations provide the possible Hilbert spa\-ces for the
system. We discuss several concrete realizations which may be
useful for future applications.

 One especially interesting
example is the space $L^2(\Real^+, r^{\alpha}\exp{(-r)}dr)$, an
orthonormal basis of which is given in terms of Laguerre's
polynomials $L_n^{\alpha}$, where $\alpha=2k-1$ in our case. As
this space is also the space of the radial wave functions for the
3-dimensional hydrogen atom, we can identify the following
correspondence: for positive integers $k$ we have $k=l+1$,
 where $l$ is
the angular momentum quantum number, and $n=n_r$, where $n_r \in
\Natural_{\, 0} \equiv \{0,1,\ldots\}$
 is the radial quantum number of the hydrogen atom!
 Thus the SBH wave functions with $k=1$ correspond to the different s-wave
 wave functions of the hydrogen atom. Because of the obvious similarities
 between  Coulomb's electrical and Newton's gravitational potentials this
 relationship may not be purely accidental.

  Section 6 discusses some (preliminary) conclusions.
Appendix A contains the main properties of the group
$SO^{\uparrow}(1,2)$, its covering groups and those features of
the irreducible unitary representations which are necessary for
our purposes. Appendix B contains the symplectic reduction of
spherically symmetric pure Einstein gravity in $D$ space-time
dimensions.
\section{Principles of group theoretical quantization}
A pedestrian way to quantize a classical system is to replace the
classical Poisson brackets of observables (functions on phase
space) by commutators of corresponding operators on a (Hilbert)
space of states. This prescription has severe limits, however: It
does not work properly for functions which are polynomials in the
basic variables $(q,p)$ of degree higher than two \cite{ste,ab}
and, another possibility, the operators may not be self-adjoint
(see, e.g. Ref. \cite{thi,ish}). Already Weyl pointed out very
early that, in order to guarantee self-adjointness of the (two)
unbounded operators $Q$ and $P$, it is advisable to pass from the
Heisenberg commutation relations $[P,Q]=\hbar/i$ etc.\ to the
bounded operators \begin{eqnarray} U(a)= e^{-iaP}\;&,&\;
 V(b)= e^{-ibQ}~~, \\
U(a)\,V(b)& =& e^{i\hbar ab} \; V(b)\,U(a)~,~ \\ U(a_1)\,U(a_2)&=&
U(a_2)\, U(a_1)~,~~ V(b_1)\, V(b_2)=V(b_2)\, V(b_1)~~
\end{eqnarray} and look for continuous irreducible unitary
representations of this (Hei\-sen\-berg-Weyl) group which provide
self-adjoint generators $P$ and $Q$. Heisenberg's commutation
relations may be interpreted as representing the Lie algebra of
 a 3-parameter group with the
group law \be (a_1,b_1,t_1) \cdot (a_2,b_2,t_2)=(a_1+a_2,b_1+b_2, t_1+t_2+
\frac{1}{2}(b_1a_2-a_1b_2))~~,\ee which represents a central extension
of the abelian (symplectic, transitive and effective)
 translation group of  $\Real^2$ interpreted as the phase
space (cotangent bundle) $T^*\Real$. According to the von
Neumann--Stone uniqueness theorem --- see, e.g., \cite{ab,thi,ish}
--- all continuous irreducible unitary representations of the
Hei\-sen\-berg--Weyl group are unitarily equivalent to the
Schroedinger representation, where the spectra of $P$ and $Q$ are
the complete real lines $\Real$.

 Group theoretical quantization
tries to generalize these properties to phase spaces (symplectic
manifolds) with nontrivial topological structures. The main steps
are (for more details we refer to the literature \cite{ish,ste};
we closely follow Isham's presentation \cite{ish}):
\begin{enumerate} \item Given a (here finite-dimensional) symplectic space
 (manifold) $\mbox{$\cal S $} = \{s\} $ with a nondegenerate symplectic
  form $\omega$, find
 a finite-dimensional Lie transformation group $\mbox{G}=\{g\}$ of
  $\cal S$ which leaves
 the symplectic form $\omega$ invariant and which acts transitively
 and effectively (i.e.\ if $ g\cdot s = s~ \forall s$, then $g=e$ (unit
 element)).
 The latter condition may be relaxed to almost effective actions (i.e.\ if
  $ g \cdot s = s~ \forall s$, then $g$ is an element of a discrete center
  subgroup). The one-parameter subgroups $g(t)= \exp(-A\,t)$ of
   $\mbox{G}$ generate
  vector fields $\tilde{A}$ on $\cal S$. As the transformations $g(t)\cdot s$
  leave $\omega$ invariant the Lie-derivatives $L_{\tilde{A}}$ have the
  property $L_{\tilde{A}}\omega=0$, which --- together with $d\omega=0$ ---
  implies $ d\,(i_{\tilde{A}}\omega)=0$, where $i_{X}$ denotes interior
  multiplication of an exterior form by a vector field $X$. The last
  relation means that $i_{\tilde{A}}\omega$ is a closed 1-form on $\cal S$.
  The corresponding vector fields $\tilde{A}$ are called
   ``locally Hamiltonian''.
  According to Poincar\'{e}'s lemma one has locally $i_{\tilde{A}}\omega = df$,
  where $f(s)$ is some function on $\cal S$. If the first cohomology group
  $H^1(\mbox{$\cal S$};\Real)$ vanishes then $i_{\tilde{A}}\omega$ is
   exact and we have a
  (globally defined) Hamiltonian vector field which --- in local canonical
  coordinates --- has the form
   \be \tilde{A} =X_f = \frac{\partial f}{\partial p_i}
  \frac{\partial}{\partial q^i}-  \frac{\partial f}{\partial q^i}
  \frac{\partial}{\partial p_i}~~. \ee
  If the vector field $X$ can be written as the commutator of two other
  vector fields, $X=[X_1,X_2]$, then, because of $i_{[X_1, X_2]}\omega =
  d(i_{X_1}i_{X_2}\omega)$, $X$ is  Hamiltonian. This is the
  case for semisimple transformation groups $\mbox{G}$.
  \item The relation (19) provides a map from  smooth functions
  $f(s)$ on $\cal S$ onto  Hamiltonian vector fields on $ \cal S$, the
  kernel of which are the constant real numbers.

  As the commutator $[X_1,X_2]$
  of two vector fields is again a vector field, the question is, which
  Hamiltonian vector field corresponds to the commutator. The answer is
  given by the Poisson bracket structure for  functions on $\cal S$:
  The Poisson bracket of two functions $f_i(s), i=1,2,$ is given by
  \be \{f_1,f_2\} = \omega(X_{f_1},X_{f_2})=-X_{f_1}(f_2) =
  \frac{\partial f_1}{\partial q^
  i}
  \frac{\partial f_2}{\partial p_i}-  \frac{\partial f_1}{\partial p_i}
  \frac{\partial f_2}{\partial q^i} \ee and we have \be [X_{f_1},X_{f_2}]=
  -X_{ \{f_1,f_2\}}~~, \ee which means that there is  an homomorphism \be f
  \rightarrow - X_f \ee from
  the Lie algebra of ``observables''  $f(s)$ onto the Lie algebra
  of Hamiltonian vector fields on $ \cal S $ with the (constant) real
  numbers as kernel. \item We now come to a crucial point of the
  quantization procedure: We have --- due to the (almost) effective action
  of the transformation group $\mbox{G}$ on $\cal S$ --- an isomorphism of the
  Lie algebra $\cal L$$(\mbox{G})$ into the Hamiltonian vector
   fields on $\cal S$ and
  a homomorphism of the Lie algebra of observables $f$ onto the Hamiltonian
  vector fields. What is needed, however, is the following: One wants an
  isomorphism between the Lie algebra $\cal L$$(\mbox{G}) = \{A\}$ and the
   Poisson algebra of a preferred set of observables $P^A(s)$ such that
   \be \tilde{A} = -X_{P^A}~, ~~~\{P^{A_1},P^{A_2}\} = P^{[A_1,A_2]}~~. \ee
Such an isomorphism --- a so-called ``momentum map'' --- is not always
possible, due to the fact that the constant functions $\in \cal S$
have vanishing Hamiltonian vector fields (Poisson brackets).  If,
however, the second cohomology group $H^2(\mbox{$\cal
L$}(\mbox{G});\Real)$ vanishes,  the momentum map does exist. This
is the case for semisimple Lie groups, like our
$SO^{\uparrow}(1,2)$.
\\ If the second cohomology group is not trivial,  one may be
forced to look for appropriate central extensions of the original
group, like in the case of the Heisenberg group, which represents
a central extension of the abelian translation group.
 \item Having established an isomorphism between the Lie algebra
  $\mbox{$\cal L$}(\mbox{G})$
 and a corresponding Poisson algebra of a system $\{P^A\}$ of preferred
 observables on $\cal S$, one then can quantize the classical system
 by using the irreducible unitary representations of the transformation
 group $\mbox{G}$ where the self-adjoint generators $K(A)$ of the
  1-parameter subgroups $U(g(t)=\exp(At))=\exp(-iK(A))$
 represent the corresponding original classical observables $P^A$.
 \item As there may be different groups with symplectic, transitive
 and effective action on $\cal S$, one has to make a choice which one to use.
 Here physical considerations come into play: One wants a group such that
 the corresponding observables $P^A(s)$ constitute basic functions on $ \cal
 S$ so that all  physically interesting observables can be expressed
 by them. For additional discussions of these problems see Ref.
 \cite{bo2}\end{enumerate}
\section{The action of the group $SO^{\uparrow}(1,2)$ on $S^1\times \Real^+$}
The local versions (3) and (13) of the symplectic form $\omega$
may belong to different global geometries and, accordingly, to
different ensuing quantum theories \cite{ish}. If we have the
usual phase space $T^*\Real \simeq \Real^2$ then the ``quantizing''
group is the abelian translation group of $\Real^2$ enlarged by a
central extension as described at the beginning of the preceding
section.

 If the phase space has the global form $T^*\Real^+ =
\{(q,p); q>0, p \in \Real\} $ the quantizing group $\mbox{G}$ is
the affine group $\mbox{G}=\{g(a,t)\,;~ a,t \in \Real\,;~
 g(a_2,t_2)\cdot g(a_1,
t_1)= g(a_2+e^{-t_2}a_1, e^{t_1+t_2})\}$ with  action $g(a,t)\cdot
 (q,p) = (e^t\,q, e^{-t}\,p -a)$. The self-adjoint generator
  $K(\mbox{S})$ of the scale
 transformations $g(0,t)$ here corresponds to the classical observable $q\,p$.

  Closer ``home'' to our system (14) is the phase space $T^*S^1=
  \{(\varphi,p)\,;~
 \varphi \in \Real \bmod 2\pi,$ $~ p\in \Real\}$ for which Isham discusses in
 detail
 the 3-parametric euclidean group $\mbox{G}\equiv
  E_2 =\{g(a_1,a_2,\theta); \theta
\in \Real \bmod 2\pi, a_1, a_2 \in \Real\}$ of $\Real^2$ with the
action $g(a_1,a_2,\theta)\cdot(\varphi,p)
=((\varphi+\theta)\;\mbox{mod } 2\pi,~ p+ a_1 \sin(\varphi
+\theta)-a_2 \cos(\varphi+\theta))$ as quantizing group. Details
can be found in Ref. \cite{ish}.

 The phase space of the last
example is still a cotangent bundle which is no longer so in our
case (14), where we have $p>0$. In that case the orthochronous
proper Lorentz group $SO^{\uparrow}(1,2)$ --- which leaves the
quadratic form $(x^0)^2-(x^1)^2 -(x^2)^2\;, x^0>0\;,$ invariant
--- appears to be the appropriate quantizing group (see also Ref.\
\cite{bo2}): The cone $ (x^0)^2-(x^1)^2 -(x^2)^2 = 0\;,~ x^0>0,$
is diffeomorphic to $\Real^2-\{0\}$:  put $x^0=p>0,
~x^1=p\cos\varphi,~ x^2= p\sin\varphi~$!

 In the following it is
advantageous to employ the twofold covering group $SU(1,1)$ of
$SO^{\uparrow}(1,2)$ (see  appendix A) the elements $g_0$ of
 which are given by
\be g_0= \left( \begin{array}{ll} \alpha & \beta \\
 \bar{\beta} & \bar{\alpha}
\end{array} \right)~,~~ |\alpha|^2-|\beta|^2=1~, \ee where $\bar{\alpha}$
means the complex conjugate of the complex number $\alpha$. If we define the
matrix \be X_0 =
    \left( \begin{array}{cc} x^0 & x^1-i\,x^2 \\ x^1+i\,x^2 & x^0
\end{array} \right)~,~~ \mbox{det}X_0= (x^0)^2- (x^1)^2-(x^2)^2~~, \ee then
the transformations $x^{\mu}\rightarrow \hat{x}^{\mu},~ \mu
=0,1,2,$ under $SO^{\uparrow}(1,2)$ are implemented by \be X_0
\rightarrow \hat{X}_0= g_0X_0g_0^+~,~~ \mbox{det}\hat{X}_0 =
\mbox{X}_0~~,\ee where $g_0^+$ denotes the hermitian conjugate of
the matrix $g_0$.

 Applying a general transformation $g_0$ to the
matrix \be \left(
\begin{array}{cc} p &  p\,e^{\ds -i\varphi} \\  p\,e^{\ds
i\varphi}
 & p
\end{array} \right)~~ \ee yields the mapping: $(p,\varphi)
\rightarrow (\hat{p}, \hat{\varphi})$, \begin{eqnarray}
 \hat{p}&=& |\alpha +e^{\ds i\varphi}\, \beta|^2 \; p~~,  \\
 e^{\ds i\hat{\varphi}}&=& \frac{\bar{\alpha}e^{\ds i\varphi} +
  \bar{\beta}}{\alpha+e^{\ds i\varphi}\;\beta}~~. \end{eqnarray}
As \be \frac{\partial \hat{\varphi}}{\partial \varphi} = |\alpha +
e^{\ds i \varphi} \beta |^{-2} \ee we have \be
d\hat{\varphi}\wedge d\hat{p} = d\varphi \wedge d p~~, \ee  that
is, the transformations (28) and (29) are symplectic.

 One sees
immediately that $g_0$ and $-g_0$ lead to the same transformations
of $p$ and $\varphi$. Thus, the group $SU(1,1)$ acts on $\cal S$
only almost effectively with kernel $\Integer_2$ representing the
center of the twofold covering group of $SO^{\uparrow}(1,2)$. It
is well-known that the latter group acts effectively and
transitively on the forward light cone and thus on $\cal S$ (see
also below).

 We next discuss the actions of the
1-parametric subgroups $\mbox{K}_0,\mbox{A}_0$ and $\mbox{N}_0$
forming the Iwasawa decomposition $SU(1,1)=\mbox{K}_0\cdot
\mbox{A}_0 \cdot \mbox{N}_0$, with the general element
\begin{eqnarray}  k_0\cdot a_0 \cdot n_0& =& \left(
\begin{array}{cc} e^{\ds -i\theta/2} & 0 \\ 0 & e^{\ds i\theta/2}
\end{array} \right)
\cdot
\left( \begin{array}{cc} \cosh(t/2) &  -i\, \sinh(t/2) \\ i\, \sinh(t/2)
& \cosh(t/2)
\end{array} \right)\nonumber
  \\  && \cdot \left( \begin{array}{cc} 1-i\xi/2 &  \xi/2 \\
\xi/2 & 1+i\,\xi/2
\end{array} \right)~,  \end{eqnarray}
where $\theta \in (-2\pi, +2\pi];~ t,\xi \in \Real$.

 The actions
of the subgroups $\mbox{K}_0,\mbox{A}_0,\mbox{N}_0$, respectively,
are the following ones:
 \begin{eqnarray} \mbox{K}_0:~&& \hat{p}=p~,\\
&&  e^{\ds i\hat{\varphi}}= e^{\ds i(\varphi+
\theta)}~~. \nonumber \\
   \mbox{A}_0:~ && \hat{p}= \rho(t,\varphi)\,p~,~~
\rho(t,\varphi)=\cosh t+\sinh t\,\sin\varphi~, \\
&&\cos\hat{\varphi}=
\cos\varphi/\rho(t,\varphi)~,~~\sin\hat{\varphi}=(\cosh
t\sin\varphi + \sinh t) /\rho(t,\varphi)~~.\nonumber \\
\mbox{N}_0:~ && \hat{p}=
\rho(\xi,\varphi)\,p~,~~\rho(\xi,\varphi)=1+\xi \cos\varphi +
\xi^2(1-\sin\varphi)/2~, \\ &&\cos\hat{\varphi}=
[\cos\varphi+\xi(1-\sin\varphi)]/\rho(\xi,\varphi)~, \nonumber \\
&&\sin\hat{\varphi}= [\sin\varphi+\xi \cos\varphi +
\xi^2(1-\sin\varphi)/2]/\rho(\xi,\varphi)~~.\nonumber
\end{eqnarray} The groups (33) and (34) act transitively on $\cal
S$: Any point $s_1=(\varphi_1,p_1)$ may be transformed into any
other point $s_2=(\varphi_2, p_2)$ in the following way: first
transform $(\varphi_1,p_1)$ into $(0,p_1)$ by
$k_0(\theta=-\varphi_1)$, then map this point into $(\hat{\varphi}
= \arctan(\sinh\hat{t}\,),p_2)$ by $a_0(\hat{t};\,
\cosh\hat{t}=p_2/p_1)$ and finally transform $(\hat{\varphi},p_2)$
by $k_0(\theta = \varphi_2-\hat{\varphi})$ into
$s_2=(\varphi_2,p_2)$. As $\mbox{K}_0$ and $\mbox{A}_0$ combined
act already transitively on $\cal S$ one might wonder whether they
alone are not sufficient for our purpose. However, they do not
form a 2-dimensional subgroup of $SU(1,1)$, only $\mbox{A}_0$ and
$\mbox{N}_0$ do. The above transitivity properties reflect the
fact that any element $g_0$ of $SU(1,1)$ may be written as
$k_0(\theta_2)\cdot a_0(t) \cdot k_0(\theta_1)$ (see
 appendix A).

The transformation formulae (35) show that the group $\mbox{N}_0$
leaves the half-line $\varphi = \pi/2,\, p>0$ invariant, that is,
$\mbox{N}_0$ is the stability group of these points. This means
that the symplectic space (14) is diffeomorphic to the coset space
$ SU(1,1)/(\Integer_2 \times \mbox{N}_0) \simeq
SO^{\uparrow}(1,2)/\mbox{N}_0$. Notice that $\mbox{N}_0$, and
$\mbox{A}_0$ as well, does not contain the second center element
$-e$ of $SU(1,1)$. The center $\Integer_2$ is a subgroup of
$\mbox{K}_0$.

 If we pass to the universal covering
group $\widetilde{SU(1,1)}$ of $SU(1,1)$ (or of
$SO^{\uparrow}(1,2)$),
 see Eq.\ (24),
\be \widetilde{SU(1,1)}=\{\tilde{g}=(\omega,\gamma);~ \omega
\equiv \arg(\alpha)
 \in \Real,~ \gamma =\beta/\alpha,\, |\gamma|<1) \}\;,\ee
 (as to the group multiplication laws see appendix A), the transformations
 (28) and (29) take the form
\begin{eqnarray}
 \hat{p}&=& \rho(\tilde{g}, \varphi)\; p~,~~\rho (\tilde{g}, \varphi)=
 |1 +e^{\ds i\varphi}\, \gamma|^2\, (1-|\gamma|^2)^{-1}~~,  \\
 e^{\ds i\hat{\varphi}}&=& e^{\ds -2i\omega}\;
 \frac{e^{\ds i\varphi} +
  \bar{\gamma}}{1+e^{\ds i\varphi}\gamma}~~. \end{eqnarray}
  As $\partial\hat{\varphi}/\partial\varphi= 1/\rho(\tilde{g},\varphi)$, the
  equality (31) holds again.

  With the elements of the group $SU(1,1)$  given by the restriction
  $ -\pi < \omega \leq +\pi,\; \alpha = \exp(i\omega)(1-|\gamma|^2)^{-1/2},\;
  \beta = \gamma\, \alpha~,$ the homomorphisms \begin{eqnarray}
   h&:&\widetilde{SU(1,1)} \rightarrow SU(1,1)~~, \\
   h_0&:& SU(1,1) \rightarrow SO^{\uparrow}(1,2)~~, \end{eqnarray}
  have the kernels
  $\mbox{ker}(h)=2\pi\,\Integer,~\mbox{ker}(h_0)=\Integer_2$,
  respectively,
   and the composite
  homomorphism $h_0 \circ h$ has the kernel $\pi \Integer$.

  As the space $\cal S$, Eq.\ (14), is diffeomorphic to $\Real^2-\{0\}
  =\Complex-\{0\}$,
  its universal covering space is given by $\varphi \in \Real,~ p \in
  \Real^+$ which is
  the infinitely sheeted Riemann surface of the logarithm. The transformations
  (37) and (38) may be interpreted as acting transitively and effectively
  on that universal covering space.

  We would like to mention that one can define genuine {\em
  effective} actions of any covering group of $SO^{\uparrow}(1,2)$
  on $ \cal S$. However, these actions violate the ``strong
  generating principle'' of Isham \cite{ish} and are not adequate
  for a group theoretical quantization \cite{bo2}.

  The action of $SO^{\uparrow}(1,2)$ on $\cal S$ may also be
  obtained as a lift of the respective subgroup of Diff($S^1$) to
  $T^*S^1 \supset \cal S$. This will be discussed further in Ref.\
  \cite{bo2}, where it is also shown that, under certain
  conditions, the group $SO^{\uparrow}(1,2)$ is unique as to the
  required action on $\cal S$.
  \section{Hamiltonian vector fields
  induced  on $\cal S$ by \\$SU(1,1)$ transformations and the
  corresponding classical observables} For infinitesimal values of the
  parameters $\theta,\, t,\, \xi$ the transformations (33)-(35) take the
  form
 \begin{eqnarray}\mbox{K}~:&& \delta \varphi = \theta,~|\theta|\ll
  1,~~~ \delta p
 =0~,\\ \mbox{A}~:&& \delta \varphi =(\cos\varphi)\,t,~~\delta p =
  p\,(\sin\varphi)\,t,~~
 |t|\ll 1~, \\ \mbox{N}~:&& \delta\varphi=(1-\sin\varphi)\,\xi,~~\delta p =
 p\,(\cos\varphi)\,\xi,~~|\xi|\ll 1~~. \end{eqnarray} They induce on $\cal S$
 the vector fields \begin{eqnarray} \tilde{A}_K &=& -\partial_{\varphi}~~,\\
 \tilde{A}_A &=&
 -\cos\varphi\,\partial_{\varphi}-p\,\sin\varphi\,\partial_p~~, \\
 \tilde{A}_N& =& (\sin\varphi
 -1)\partial_{\varphi}-p\,\cos\varphi\,\partial_p~~. \end{eqnarray}
 It follows from the general considerations of the preceding section and it
 is easy to check that their Lie algebra is isomorphic to the
 Lie algebra of $SO^{\uparrow}(1,2)$ (see appendix A)
 (and all its covering groups, of course). A general element is the
 linear combination \begin{eqnarray}
  \tilde{A}(\lambda_K,\lambda_A, \lambda_N)& =&
\lambda_K \tilde{A}_K+\lambda_A \tilde{A}_A+ \lambda_N\tilde{A}_N
\\ &=& -(\lambda_A\, p\,
\sin\varphi+\lambda_N\,p\,\cos\varphi)\partial_p \nonumber \\
&&-(\lambda_K +\lambda_A\,\cos\varphi+\lambda_N\,
(1-\sin\varphi))\partial_{\varphi}~~. \nonumber \end{eqnarray} One
sees immediately that this vector field can be identified with the
Hamiltonian vector field \begin{eqnarray} -X_f&=&\frac{\partial
f}{\partial \varphi}\partial_p - \frac{\partial f}{\partial
p}\partial_{\varphi}~~, \\
f(\varphi,p)&=&\lambda_K\,p+\lambda_A\,p\,
\cos\varphi+\lambda_N\,p\,(1-\sin\varphi)~~. \end{eqnarray} If we
replace the Lie algebra element $l_{N}$ by $l_{B}=l_{N}-l_{K}$
(see appendix A), then the observable $f$ becomes \be
f(\varphi,p)= \lambda_K\,p+\lambda_A\,p\,
\cos\varphi-\lambda_B\,p\,\sin\varphi~~, \ee and we see that the
associated 3 basic classical observables are \be P^K =p,~~ P^A=
p\,\cos\varphi,~~P^B = - p\,\sin\varphi~~. \ee As any smooth
function $f(\varphi,p)$ periodic in $\varphi$ with period $2\pi$
can, under certain conditions, be
 expanded in a Fourier series and as $\sin(n\varphi)$ and
$\cos(n\varphi)$ can be expressed as polynomials of $n$-th order
in $\sin\varphi$ and $\cos\varphi$, the observables (51) are
indeed basic ones on $\cal S$. Actually they are just the
Cartesian coordinates of $\Real -\{0\}$ we started  with, see Eq.
(27).
\section{Quantum spectrum of the area operator \\ and associated
 Hilbert spaces}
We now come to the quantization of the classical system we have
been discussing. It consists in replacing each of the three basic
observables (51) by the self-adjoint generator $K(l)$ of the
unitary operator $U_l(t)=\exp(-i\,K(l)\,t)$ (or
$\exp(i\,K(l)\,t)$),
 representing any of the associated
 1-parameter subgroups $\exp(l\,t),l\in \mbox{$\cal L$}SO^{\uparrow}(1,2)$,
in an appropriate irreducible unitary representation of
$SO^{\uparrow}(1,2)$ or its covering groups: Thus, the observable
$p$ is to be replaced by the self-adjoint generator $K_3 \equiv
K(l_K)$ of the unitary operator $U(\theta) = \exp(-iK_3\, \theta)$
representing the compact subgroup
$\mbox{K}=\{\exp(l_{K}\,\theta)\} $ and the observables
$p\,\cos\varphi$ and $-p\, \sin\varphi$ are to be replaced by the
corresponding self-adjoint generators $K_1$ and $K_2$ of the
unitary operators $U(t_1)$ and $U(t_2)$
 representing the (``boost'') subgroups $\mbox{A}$ and $\mbox{B}$.

  In this section we
 mainly use the known properties of the irreducible unitary representations of
$SO^{\uparrow}(1,2)$ and $\widetilde{SO^{\uparrow}(1,2)}$. More
details and  references to the literature are contained in
appendix A.

 We first put $\hbar =1$ and restore it explicitely
later.

 Because $K_3$ is associated with a compact subgroup, its
spectrum is discrete in all irreducible unitary representations of
$SO^{\uparrow}(1,2)$ or its covering groups. However, not all
irreducible unitary
 representations  are suitable for our purposes, because $\hat{p}= K_3$
  corresponds to a classical area, Eq.\ (12), which is positive. Thus,
  we are interested in those irreducible representations for which the
  spectrum of $K_3$ is positive. This is the case for the so-called
   ``positive
  discrete series'' of irreducible unitary representations.
  These representations are characterized by the value $k(1-k)$ of the
  associated Casimir operator $Q=K_1^2+K_2^2 -K_3^2$, where $k$ can take
  the values $1,2,\ldots, $ for a ``true''
  representation of $SO^{\uparrow}(1,2)$,
  but $k$ can be any positive real number $> 0 $ for the
   corresponding representations
  of the universal covering group
  $\widetilde{SO^{\uparrow}(1,2)}$. For the groups $SU(1,1)
  \simeq SL(2,\Real)$
  the number $k$ can take the values $1/2,1,3/2,2, \ldots $. In all
   cases the operator $\hat{p}=K_3$ has the
  spectrum \be \mbox{spec}\,(\hat{p} \equiv K_3)= \{ k+n,\,
   n \in \Natural_{\, 0}\}~~.
  \ee
 For the representations of the universal covering group
  $k \bmod 1 $ represents the so-called
  ``$\theta$-parameter'' which occurs in other unitary representations
  involving the infinitely sheeted compact group $SO(2)$ \cite{ish,ree}.
  For a more general setting of that parameter in connection with
   multiply connected
  symplectic manifolds see again Ref.\
  \cite{ish}.

  As only $SO^{\uparrow}(1,2)$ acts effectively on the symplectic space
  $\cal S$ of Eq.\ (14), the $\theta$-parameter comes  into play merely
  if we allow for almost effective group actions by the
   universal covering group.
   Whether one has to do so
   or not, finally has to be decided by physical
  considerations.

  We next come to the discussion of concrete Hilbert spaces on which
  the operators $K_i$, $i=1,2,3,$ act as self-adjoint operators and where
  $K_3$ has one of the spectra (52).
\subsection{Hilbert space of holomorphic functions \\ inside the unit
disc $\cal D$}
   Probably the most important Hilbert space is
  the (Bargmann) Hilbert space $ \mbox{$\cal H$}_{\mbox{$\cal D$},\, k}$
   of holomorphic functions in the unit
  disc $\mbox{$\cal D$}=\{z=x+iy,\, |z| <1 \}$ with the scalar product
  \be (f,g)_{\mbox{$\cal D$},\, k}
   =\frac{2k-1}{\pi}\int_{\mbox{$\cal D$}} \bar{f}(z)g(z)\,(1-|z|^2)^{2k-2}
  dxdy~~. \ee It can be used for any real $k > 1/2$ and also in the
  limiting case $k\rightarrow 1/2$. As
  \be  (z^{n_1},z^{n_2})_{\mbox{$\cal D$},\, k}
   = \frac{\Gamma(2k)\,\Gamma(n_1+1)}{\Gamma(2k+n_1)}
 \; \delta_{n_1\,n_2}~, \ee and since any holomorphic
  function in $\mbox{$\cal D$}$ can be
  expanded in powers of $z$ the functions
  \be e_{k,n}(z)= \sqrt{\frac{\Gamma(2k+n)}{\Gamma(2k)\,\Gamma(n+1)}}\,
  z^n~~,~~ n \in \Natural_{\, 0}~~, \ee form an orthonormal basis of
  $\mbox{$ \cal H$}_{\mbox{$\cal D$},\, k}$.
 The operators $K_i$ here have the explicit forms
 \begin{eqnarray}  K_3 &=&k+z\frac{d}{dz}~~,
  \\ K_+&\equiv &K_1+i\,K_2 =2kz+z^2\frac{d}{dz}~~, \\
 K_-&\equiv &K_1-iK_2=\frac{d}{dz}~~. \end{eqnarray} The basis functions (55)
 are the eigenfunctions of $K_3$ with eigenvalues $k+n$, the operators $K_+$
 and $K_-$ being raising and lowering operators:
 \begin{eqnarray} K_3\,e_{k,n}&=&(k+n)\,e_{k,n}~~, \\
 K_+\,e_{k,n}&=&[(2k+n)(n+1)]^{1/2}\,e_{k,n+1}~~, \\
 K_-\, e_{k,n}&=&[(2k+n-1)n]^{1/2}\, e_{k,n-1}~~. \end{eqnarray}
The formulae (1) and (9) suggest to associate them with the
irreducible representation $k=1$, that is with scalar product and
eigenfunctions \be (f,g)=\frac{1}{\pi}\int_{\mbox{$\cal
D$}}dxdy\bar{f}(z)g(z)~,~~e_{1,n}=\sqrt{n+1}\; z^n,~n \in
\Natural_{\, 0}~. \ee If we have on $ \mbox{$\cal D$}$ the
holomorphic functions \be
f(z)=\sum_{n=0}^{\infty}a_n\,z^n~,~~g(z)=
\sum_{n=0}^{\infty}b_n\,z^n~, \ee then, according to Eq.\ (54),
 their scalar product $(f,g)_{\mbox{$\cal D$},\, k}$ is given by
\be (f,g)_{\mbox{$\cal D$},\, k}=\sum_{n=0}^{\infty}
\frac{\Gamma(2k)\,\Gamma(n+1)}{\Gamma(2k+n)}
 \bar{a}_n \,b_n~~.\ee
This series can be used as a scalar product to extend the
definition of the Hilbert spaces $\mbox{$\cal H$}_{ \mbox{$\cal
D$},\, k}$ to all real $k>0$! \subsection{The Hardy space of the
unit circle} For the special case $k=1/2$ the coefficient in front
of $\bar{a}_n\, b_n$ in (64) has the value 1. This allows for a
reinterpretation of the Hilbert space $ \mbox{$\cal
H$}_{\mbox{$\cal D$},\frac{1}{2}}$: Consider the $L^2$-space on
the unit circle with the scalar product \be
(\psi_1,\psi_2)=\frac{1}{2\pi}\int_0^{2\pi}d\phi \;
\bar{\psi}_1(\phi)\,\psi_2(\phi)~~,\ee an orthonormal
 basis of which is given by the
functions $\exp(in \phi)\;,~n \in \Integer$. \\That subspace of
functions $h(\phi) \in  L^2$ which have only ``positive'' Fourier
coefficients, $ a_n =0,\; n<0$,
 is called the ``Hardy space $H^2_+$ of the unit circle'',
  and the corresponding scalar product
 is denoted by $(h_1,h_2)_+$. It has the orthonormal
 basis $\exp(in\phi)\,,~n \in \Natural_{\, 0}~.$

 Hardy spaces \cite{ho,ru,con} have a number of interesting
 properties and are closely related to Hilbert spaces of holomorphic
 functions \cite{bar2,ho,ru,con}.

  If we have the two Fourier series $\in H^2_+$
\be h_1(\phi)=\sum_{n=0}^{\infty}a_n\,e^{\ds in\phi}~,~
~h_2(\phi)= \sum_{n=0}^{\infty}b_n\,e^{\ds in\phi}~~, \ee they
have the scalar product \be (h_1,h_2)_+=
\frac{1}{2\pi}\int_0^{2\pi} d\phi\;
\bar{h}_1(\phi)h_2(\phi)=\sum_{n=0}^{\infty}\bar{a}_n \,b_n~~. \ee
Thus we may realize the Hilbert space $ \mbox{$\cal
H$}_{\mbox{$\cal D$},\frac{1}{2}}$ by using the Hardy space
$H^2_+$!
\subsection{Unitary representations on Hardy space related Hilbert spaces}
 What is the relation of the other spaces $
\mbox{$\cal H$}_{\mbox{$\cal D$},\, k}$ to the Hardy space
$H^2_+$? The answer is somewhat subtle \cite{puk,sa2,vil}:
 Define the self-adjoint operator $A_k$ in $H^2_+$ which is diagonal
in the basis $\{\exp(in\phi)\}$ of $H^2_+$ and which acts on it as
\be A_k\,e^{\ds in\phi}=
\frac{\Gamma(2k)\,\Gamma(n+1)}{\Gamma(2k+n)}\; e^{\ds in\phi}~~.
\ee Then define an $H^2_+$ related Hilbert space $H^2_{A_{k}}$
with the scalar product \be (h_1,h_2)_k = (h_1,A_k\,h_2)=
\sum_{n=0}^{\infty} \frac{\Gamma(2k)\,\Gamma(n+1)}{\Gamma(2k+n)}
 \bar{a}_n\,b_n~~  \ee for the functions (66). The series (69)
 representing the scalar product of $H^2_{A_k}$ is obviously
 the same as the series (64) which represents the scalar product for
$\mbox{$\cal H$}_{\mbox{$\cal D$},\, k}$. This exhibits the very
close relationship between the two Hilbert spaces. The
mathematical background for this is that holomorphic functions
inside the unit disc $\mbox{$\cal D$}$ have holomorphic limits on
$\partial \mbox{$\cal D$}=S^1$ (for mathematical details see the
Refs. \cite{ho,ru,con}).
\\ An orthonormal basis for $H^2_{A_k}$ is given by
\begin{eqnarray}
 \chi_{k,n}(\phi)&=& \sqrt{\frac{\Gamma(2k+n)}{\Gamma(2k)\,\Gamma(n+1)}}\,
  e^{\ds i(k+n)\phi},\, n \in \Natural_{\, 0}~, \\&& ~~~~~~~~~~~~~~~~~~~~~~
  (\chi_{k,n_1},\chi_{k,n_2})_k =
  \delta_{n_1n_2}~~,\nonumber
   \end{eqnarray}  where we have included an
  overall phase factor $\exp(ik\phi)$.

   With respect to this basis
  the  operators $K_3,\,K_+,\,K_-$ have the form
  \be K_3=\frac{1}{i}\,\partial_{\phi}~,~~K_+=e^{\ds
  i\phi}(ik+\partial_{\phi})~,~~K_-= e^{\ds -i\phi}(ik-\partial_{\phi})~~.
  \ee Their action on the basis functions (70) is given by
  \begin{eqnarray} K_3\chi_{k,n}&=&(k+n)\chi_{k,n}~~, \\
 K_+\chi_{k,n}&=& i[(2k+n)(n+1)]^{1/2}\chi_{k,n+1}~~, \\
 K_-\chi_{k,n}&=& \frac{1}{i}[(2k+n-1)n)]^{1/2}\chi_{k,n-1}~~.
\end{eqnarray}
It is important to realize that the  operators $K_3,\, K_+,\, K_-$
belong to a  representation which is unitary only with respect to
the scalar product (69), not with respect to the scalar product
(67)! This may be seen explicitly as follows: Applying the
operators $K_+$ and $K_-$ from Eq.\ (71) to the series \be
f_1(\phi)=\sum_{m=0}^{\infty}a_m\,\chi_{k,m}(\phi)~,~ ~f_2(\phi)=
\sum_{n=0}^{\infty}b_n\,\chi_{k,n}(\phi)~~,
 \ee using the relations (73) and (74) and the orthonormality
 (70) yields \be (f_2,K_+f_1)_k = \sum_{n=0}^{\infty}
 i [(2k+n)(n+1)]^{1/2}\,\bar{b}_{n+1}\,a_n = (K_-f_2,f_1)_k~~, \ee
 which says that $K_-$ is the adjoint operator of $K_+$ with respect
 to the scalar product (69). But one sees immediately that this is not so
 with respect to the scalar product (67)!

  Furthermore, the multiplication
 operator $\exp(i\phi)$ is not a unitary operator on  $H_+^2$, because its
 inverse does not always exist: for instance, the constant function
 $f=1$
 is an element of $H_+^2$, but $\exp(-i\phi)\cdot 1= \exp(-i\phi)$ is not.
 Such isometric operators are called ``shift operators''  and their
 properties have been investigated systematically by the mathematicians
 \cite{nag,ho,ru,con}.

The question is, whether there are irreducible unitary
representations of the positive discrete series of
$SO^{\uparrow}(1,2)$ or its covering groups on the Hardy space
$H_+^2$ itself? Sally has shown \cite{sa1}, by a detour, that
there are such representations on $H_+^2$ and that they are
unitarily equivalent to the ones above. We shall briefly indicate
how this works, because on the way we learn about other
interesting Hilbert spaces on
 which some of the above irreducible representations
are realized. \subsection{Unitary representations in the Hilbert
space \\ of  holomorphic functions on the upper half plane} The
unit disc $\mbox{$\cal D$}$ and its associated Hilbert space with
 the scalar product (53)
is especially suited for the construction of unitary
representations of $SU(1,1)$ because that group acts transitively
on $\mbox{$\cal D$}$ (see appendix A). Similarly, the group
$SL(2,\Real)$, which is isomorphic to $SU(1,1)$, see appendix A,
acts transitively on the upper complex half plane
$\Complex^{+i}=\{w=u+iv,\; v>0\}$. The mapping \begin{eqnarray}
w&=&\frac{1-iz}{z-i}=
\frac{2x+(1-x^2-y^2)\,i}{x^2+(y-1)^2}~,\\z&=&\frac{iw+1}{w+i}~,~~|z|^2
=\frac{u^2 +(v-1)^2}{u^2+(v+1)^2}~~, \end{eqnarray} provides a
holomorphic diffeomeorphism from $\mbox{$\cal D$}$ onto
$\Complex^{+i}$ and back. Because of \be \frac{dudv}{4v^2}
=\frac{dxdy}{(1-|z|^2)^2}~,~~1-|z|^2 =
\frac{2^2\,v}{(w+i)(\bar{w}-i)} ~~,\ee we have for $k=1/2,1,3/2,2,
\ldots $ the following isomorphism: \be (f,g)_{\mbox{$\cal D$},\,
k}=(\tilde{f},\tilde{g})_{\Complex^{+i},\,
k}=\frac{1}{\Gamma(2k-1)}
 \int_{\Complex^{+i}}\overline{\tilde{f}}~\tilde{g}\, v^{2k-2}dudv~~,\ee
 where \begin{eqnarray} E_k:&& \tilde{f}(w)=\sqrt{\frac{\Gamma(2k)}{\pi}}
\, 2^{2k-1}(w+i)^{-2k} f\left(z=\frac{1+iw}{i+w}\right)~~,\\
E^{-1}_k:&& f(z)= 2\sqrt{\frac{\pi}{\Gamma(2k)}}\,
(z-i)^{-2k}\tilde{f}\left(w=
\frac{1-iz}{z-i}\right)~~. \end{eqnarray}
The  (unitary) transformation $E_k$ maps the basis (55) of
 $ \mbox{$\cal H$}_{\mbox{$\cal D$},\, k}$ onto the basis
 \be \tilde{e}_{k,n}(w)= \sqrt{\frac{\Gamma(2k+n)}{\pi\Gamma(n+1)}}
\;2^{2k-1}\;i^n\;(w-i)^n\;(w+i)^{-2k-n}~,n \in \Natural_{\, 0}~,
\ee of $ \mbox{$\cal H$}_{\Complex^{+i},\, k}$.\\ One can, of
course, discard the phase factor $i^n$. \\ On this Hilbert space
 the irreducible unitary
representations $T_k^+$ of the positive discrete series of
$SL(2,\Real)$ are given by \begin{eqnarray}
[T^+(g_1,k)\tilde{f}](w)& =&
(a_{12}w+a_{22})^{-2k}\tilde{f}\left(\frac{a_{11}w+a_{21}}{a_{12}
w+a_{22}}\right)~, \\ g_1& =& \left( \begin{array}{cc} a_{11}&
a_{12}\\ a_{21}&a_{22} \end{array}\right) \in SL(2,\Real)~,
\end{eqnarray} which is defined for $k=1/2,1,3/2,2, \ldots $ only.
The subgroups
\begin{eqnarray} \mbox{K}_1: && k_1 = \left( \begin{array}{cc} \cos(\theta/2)&
\sin(\theta/2)\\ -\sin(\theta/2)& \cos(\theta/2) \end{array}
\right)~~,\\ \mbox{A}_1:&& a_1 = \left(
 \begin{array}{cc} e^{\ds t_1/2}&
  0\\ 0 & e^{\ds -t_1/2} \end{array} \right)~~, \\ \mbox{B}_1:&&
  b_1 =
\left( \begin{array}{cc} \cosh(t_2/2)&
\sinh(t_2/2)\\ \sinh(t_2/2)& \cosh(t_2/2) \end{array} \right)~~
\end{eqnarray}
are associated with the following generators of their unitary
representations (we choose the sign of $\tilde{K_3}$ such that its
spectrum is positive): \begin{eqnarray} \tilde{K}_3&=
&\frac{1}{i}( k\,w+\frac{1}{2}(w^2+1)\frac{d}{dw})~, \\
\tilde{K}_{\pm}&=& \pm k\,(w\mp i)\pm \frac{1}{2}(w\mp
i)^2\frac{d}{dw}~.
\end{eqnarray} Their action on the basis (83) is given by
 \begin{eqnarray} \tilde{K}_3\tilde{e}_{k,n}&=&(k+n)\tilde{e}_{k,n}~~, \\
\tilde{K}_+\tilde{e}_{k,n}&=& i[(2k+n)(n+1)]^{1/2}\tilde{e}_{k,n+1}~~, \\
\tilde{K}_-\tilde{e}_{k,n}&=& \frac{1}{i}[(2k+n
-1)n)]^{1/2}\tilde{e}_{k,n-1}~~.
\end{eqnarray}
For the limiting case $k \rightarrow 1/2$ the Hilbert space with
the scalar product (80) now can be replaced by the ``Hardy space
$H^2_{+i}$ of the upper half plane'', \cite{ho,ru,con} the elements
of which are the functions $\tilde{f}(u)$ which are limits for
$\Im(w)=v \rightarrow 0$ of the
 previous holomorphic functions $\tilde{f}(w)$
  on the upper half plane and the Hilbert
 space of which  has the scalar product \be (\tilde{f_1},\tilde{f_2})_{+i}
 =\int_{-\infty}^{\infty}du \overline{\tilde{f}}_1(u)\,\tilde{f}_2(u)~~.\ee
 \subsection{Hilbert space of the Fourier transformed \\ holomorphic functions
 on the upper half plane}
 We now pass to still another Hilbert space  $ \widehat{\mbox{$\cal
  H$}}_{\Complex^{+i},\, k}$ by the Fourier transform \be
  \mbox{$\cal F$}:~~~~~~
   \hat{f}(t)
  = \frac{1}{\sqrt{2\pi}} \int_{-\infty}^{\infty}\tilde{f}(w)e^{\ds
  -itw}du~,~~\tilde{f}(w) \in \mbox{$\cal
  H$}_{\Complex^{+i},\, k}~,~~t\in \Real~. \ee
  Because of the analyticity properties of $\tilde{f}$
  one has \cite{ru2} $\partial_v \hat{f}(t)=0$ and
    $\hat{f}(t) = 0$ for $t<0$ and the inversion is given by
  \be \mbox{$ \cal F$}^{-1}:~~~~~~~~~~~~\tilde{f}(w)=\frac{1}{\sqrt{2\pi}}
  \int_0^{\infty}\hat{f}(t)e^{\ds iwt}dt~~. \ee The scalar product induced on
 $ \widehat{\mbox{$\cal H$}}_{\Complex^{+i},\, k}$ is  \be
 (\tilde{f},\tilde{g})_k=(\hat{f},\hat{g})_k
 =\frac{1}{2^{2k-1}}\int_0^{\infty}\overline{\hat{f}}(t)\hat{g}(t)t^{1-2k}dt~.
 \ee The Fourier transform (95) maps the basis (83) of $ \mbox{$\cal
  H$}_{\Complex^{+i}, \, k}$ onto the basis \cite{it1}
  \be \hat{e}_{k,n}(t)=i^{n-2k}\sqrt{\frac{2\Gamma(n+1)}{\Gamma(2k+n)}}
  (2t)^{2k-1}e^{-t}L^{2k-1}_n(2t) \ee of $ \widehat{\mbox{$\cal
  H$}}_{\Complex^{+i},\, k}$. Here $L^{2k-1}_n$ are Laguerre's polynomials
  which obey the equation \cite{htf2} \be
   x\,L^{2k-1\;''}_n +(2k-x)L^{2k-1\;'}_n +n\,L^{2k-1}_n
  =0~.\ee Using the inverse Fourier transform (96) the
   operator $\tilde{K}_3$
  can be seen to  take now
  the form \be \hat{K}_3 = -\frac{1}{2}
  \frac{d^2}{dt^2}t + k\frac{d}{dt}+\frac{t}{2}~,\ee of
   which the basis functions (98) are eigenfunctions
  with eigenvalue $k+n$. \\
  It is  important  that on $ \widehat{\mbox{$\cal
   H$}}_{\Complex^{+i},\, k}$ the
  parameter $k$ can take any value $>0$,
  contrary to $ \mbox{$\cal
  H$}_{\Complex^{+i},\,k}$ where $k$ can take only the
   values $1/2,1,3/2,2, \ldots$.
\subsection{Unitary representations on the Hilbert space $L^2(\Real^+, dt)$}
The measure $dt/(2t)^{2k-1}$ in the scalar product (97) and
 the form of the eigenfunctions
(98) strongly suggest to introduce the unitary mapping \be
V_k:~~~~~~~~~~~~~~~~~ \hat{f}(t) \rightarrow \check{f}(t)=
 \hat{f}(t)(2t)^{1/2-k} \ee of $L^2(\Real^+,(2t)^{1-2k}\,dt)$ onto
 $ \widehat{\mbox{$\cal  H$}}_{+,\, k}$, the standard Hilbert space
 $L^2(\Real^+,dt)$
  on the positive real line with the standard orthonormal basis
 \be f_{k,n}(t)=\sqrt{\frac{2\Gamma(n+1)}{\Gamma(2k+n)}}
  (2t)^{k-1/2}e^{-t}L^{2k-1}_n(2t) = i^{2k-n}\check{e}_{k,n}(t)~. \ee
The operator $\check{K}_3$ here has the form \be \check{K}_3 = -\frac{1}{2}(
t\frac{d^2}{dt^2}+\frac{d}{dt})+\frac{1}{2}t + \frac{(2k-1)^2}{8t}~, \ee
with the property \be \check{K}_3 f_{k,n}(t)=(k+n)f_{k,n}(t)~. \ee
\subsection{Relationship to the radial wave functions \\ of the
hydrogen atom} If we define $f_{k,n}=t^{1/2}h_{k,n}(t)$, then the
eigenvalue equation (104) can be rewritten as \be
-\frac{1}{2}\left(h^{''}_{k,n}+\frac{2}{t} h^{'}_{k,n}(t)\right)
+\left(\frac{k(k-1)}{2t^2}-\frac{k+n}{t}\right)h_{k,n}(t)=
-\frac{1}{2}h_{k,n}(t)~. \ee This is just the radial
Schr\"{o}dinger Eq.\ for the hydrogen atom in 3 space dimensions
with mass $m=1$, angular momentum $k= l+1$, fine structure
constant $\alpha = k+n$ and bound state energy $E_{l,\,n_r}=
-1/2$. As \be E_{l,\, n_r}=-\frac{1}{2}
\frac{\alpha^2}{(l+n_r+1)^2}~\ee for the energy levels of the
hydrogen atom, we see that our quantum number $n$ here is to be
identified with the radial quantum number $n_r=0,1,\ldots$.

 Thus, for $k=1,2, \ldots~,$
 we have related the
quantum theory of the Schwarz\-schild black hole to that of the
hydrogen atom with varying fine structure constant. The
irreducible representation with $k=1$ corresponds to the s-wave
bound states of the hydrogen atom.

 As in the gravitational case
$\alpha \propto M^2$, we see that we are consistent here with the
relation (2). \subsection{Unitary representations on the Hardy
spaces \\ of the upper half plane and the unit circle} Next we
give the form of the eigenfunctions of $K_3$ in the Hardy spaces
$H^2_{+i}$ and $H^2_+$ with the scalar products (94) and (67)
respectively:
\\ We have
 to apply the inverse Fourier transformation (96) to the functions
 $\check{f}_{k,n}(t)$ with real $w=u$ and in this context use the relations
  \cite{ob} \begin{eqnarray}
&& \int_0^{\infty}e^{\ds -pt}t^{ k-1/2}L_n^{ 2k-1}(t)dt = \nonumber
  \\ &=& \frac{\Gamma(2k+n)
 \Gamma (k+1/2)}{\Gamma(n+1)\Gamma(2k)}p^{-k-1/2}F\left(-n,k+
 \frac{1}{2};2k;\frac{1}{p}\right)
\\ &=& \frac{\Gamma(k+n+1/2)}{\Gamma(n+1)}(p-1)^n p^{-n-k-1/2} \times \\
&& \times F\left(-n,k-
\frac{1}{2};
 \frac{1}{2}-k-n;\frac{p}{p-1}\right)~, \nonumber \\ &&
  p=\frac{1}{2}(1-iu)~,\nonumber
  \end{eqnarray} where \begin{eqnarray} F(a,b;c;z)&=&
  1+\frac{ab}{c}z+\frac{a(a+1)b(b+1)}{c(c+1)} \frac{z^2}{2!}+\cdots \\ &&
  +\frac{a(a+1)\cdots (a+\nu-1)b(b+1)\cdots (b+\nu-1)}{c(c+1)\cdots
  (c+\nu-1)}\frac{z^{\nu}}{\nu!} \cdots \nonumber \end{eqnarray}
  is the hypergeometric series \cite{htf11} which here is a polynomial
  of degree $n$ because $a=-n$. Since \cite{htf12} \be \Gamma(k+\frac{1}{2}
  )=\frac{
  \sqrt{\pi}\;\Gamma(2k)}{2^{2k-1}\Gamma(k)}~, \ee we get  the orthonormal
  system of eigenfunctions $\tilde{f}_{k,n}(u)$ of $K_3$ on $H^2_{+i}$:
  \begin{eqnarray} \tilde{f}_{k,n}(u)&=& 2^{-k+1/2}\sqrt{
  \frac{\Gamma(2k+n)}{\Gamma(n+1)}}
  \frac{\Gamma(2k)}{\Gamma(k)}(1-iu)^{-k-1/2} \times \nonumber\\ &&
   \times F\left(-n, k+\frac{1}{2};
   2k;\frac{2}{1-iu}\right)
  ~, \\ && (1-iu)^k= (1+u^2)^{k/2}e^{ik\phi},-\frac{\pi}{2} <
  \phi= -\arctan(u)<+\frac{\pi}{2}~.\nonumber \end{eqnarray}
  This set of orthonormal functions on $H^2_{+i}=L^2(\Real,du)$
   can be interpreted
  in the framework of orthogonal polynomials \cite{sz} in
   the following manner:
  For any $k>0$ define the weight function \be \tilde{w}_k(u) = \frac{\Gamma^2
  (2k)}{2^{2k-1} \Gamma^2(k)} (1+u^2)^{k-1/2} \ee and the polynomials
  of degree $n$: \be \tilde{b}_{k,n}(u)=
  \sqrt{\frac{\Gamma(2k+n}{\Gamma(n+1)}}~ F\left(-n,k+
  \frac{1}{2};2k;\frac{2}{1-iu}\right)~. \ee
  Then the scalar product (94) may written as
  \be (\tilde{f}_{k,n_1},\tilde{f}_{k,n_2})_{+i}
  =\int_{-\infty}^{\infty} du\;
  \tilde{w}_k(u)\;\overline{\tilde{b}}_{k,n_1}(u)\; \tilde{b}_{k,n_2}(u)
  =\delta_{n_1n_2}~. \ee The operator $K_3$ now is no longer a pure
  differential operator, but due to the term $\propto t^{-1}$ in Eq.\ (103),
  an integro-differential operator on $H^2_{+i}$. \\
   In order to get the eigenfunctions $f_{k,n}(\phi)$
   of $K_3$ in $H^2_+$ on the unit circle
   we have to follow up the inverse Fourier transformation (108)
    from above by
   the mapping --- see Eq. (82) --- \be E_{1/2}^{-1}: f(z)=2
    \sqrt{\pi}\;(z-i)^{-1}
   \tilde{f}\left( w=u=\frac{1-iz}{z-i}\right)~,~|z|=1~. \ee Observing
   that $p/(p-1)=-1/(iz)$ and using the relations
   \cite{htf11}
   \be z^{-a} F(a,a-c+1;a-b+1;\frac{1}{z})=F(a,b;c;z) \ee and (108),
    we then finally get for $k \geq 1/2$ \begin{eqnarray}
f_{k,n}(\phi)&=& \gamma_{k,n}\,
(1-e^{i\phi})^{k-1/2}\;F\left(-n,k+\frac{1}{2};\frac{3}{2}-k-n;e^{i\phi}
\right)~, \\
&& \gamma_{k,n}=i \frac{\Gamma(n+k+1/2)}{\sqrt{\Gamma(n+1)\Gamma(2k+n)}}~,
\nonumber
 \end{eqnarray}
where we have put $-iz=\exp(i\phi)$ because $|z|=1$. \\ We now may proceed as
before: As \[(1-e^{i\phi})(1-e^{-i\phi})=2(1-\cos\phi)=4\sin^2(\phi/2)\]
 we
may define the weight \be w_k(\phi)= 2^{2k-1}\sin^{2k-1}(\phi/2) \ee and the
orthogonal polynomials \be b_{k,n}(\phi)=
\gamma_{k,n}\;
F\left(-n,k+\frac{1}{2};\frac{3}{2}-k-n;e^{i\phi}\right) \ee
 for all $k>0$ and can then write the scalar
product (67) as \be (f_{k,n_1},f_{k,n_2})_+=\frac{1}{2\pi}
\int_0^{2\pi}d\phi\;
w_k(\phi)\;\overline{b}_{k,n_1}(\phi)\;b_{k,n_2}(\phi)=
\delta_{n_1n_2}~. \ee We repeat the basic difference between the
eigenfunctions (70) and (117) of the self-adjoint generator $K_3$
of the corresponding unitary representations of the compact
subgroup of $SO^{\uparrow}(1,2)$ both sets of which belong to the
same vector space: The set (70) belongs to the representations
which are unitary with respect to the scalar product (69) whereas
the set (117) is associated with the scalar product (67). Both
representations are unitarily equivalent: this follows from the
sequence of mappings we have been using and which are all unitary.
\subsection{Other unitary representations  on the \\ Hardy space
$H^2_+$ of the unit circle} One can implement the constituting
relations (A.60)--(A.62) for a unitary representation on the Hardy
space $H^2_+$ with the scalar product (67) and the basis
$\exp(i\,n\,\phi),~n \in \Natural_{\, 0}$, by chosing for the
``ladder'' operators $K_{\pm}$ ``nonlocal'' expressions
\cite{bo2}:
\begin{eqnarray} \breve{K}_3 &=& k+\frac{1}{i}\frac{d}{d\phi}~,\\
\breve{K}_+&=& e^{\ds i\,\phi}\left [\left (2k+
\frac{1}{i}\frac{d}{d\phi}\right)\left(1+\frac{1}{i}\frac{d}{d\phi}
\right)\right
]^{1/2}~,~ \breve{K}_-=(\breve{K}_+)^+ ~,\end{eqnarray} because
\be \breve{K}_+ e^{i\, n\, \phi}=
[(2k+n)(n+1)]^{1/2}\,e^{i\,(n+1)\,\phi}~. \ee
\subsection{Unitary representations in the \\ state space of
 two harmonic oscillators}
Finally we mention that all irreducible unitary representations of
the positive discrete series  of $SU(1,1)$ with $k=1/2,1,3/2,2,
\ldots$ are contained in the tensor product of the Hilbert spaces
of two harmonic oscillators \cite{bi,per,how}, generated by
creation and annihilation operators \be [a_i,
a_j^+]=\delta_{ij}~,~~[a_i,a_j]=0~,~~[a_i^+,a_j^+]=0~,~~i,j=1,2~.
\ee The operators \be K_3=\frac{1}{2}(a_1^+a_1+a_2^+a_2+1)~,~~
K_+=a_1^+a_2^+~, ~~K_-=a_1a_2 \ee obey the commutation relations
\be [K_3,K_+]=K_+~,~~[K_3,K_-]=-K_-~,~~[K_+,K_-]=-2K_3 \ee of the
Lie algebra of $SO^{\uparrow}(1,2)$ and its covering groups. The
``ground state'' $|k;0\rangle$ is defined by the property \be
a_j|k;0\rangle=0~,~~j=1,2, \ee and the other normalized states by
\be |k;n_1,n_2\rangle=
\frac{1}{\sqrt{n_1!\,n_2!}}(a_2^+)^{n_2}\,(a_1^+)^{n_1}
|k;0\rangle~,~~ n_j\in \Natural_{\, 0}~. \ee Notice that $K_3$ is
just half the sum of the two Hamilton operators
$H_j=(a_j^+a_j+1/2), j=1,2,$ of the two harmonic oscillators.

 The relation between the
number pair $(n_1,n_2)$ and the pair $(k,n),n \in \Natural_{\,
0},$ characterizing a state in an irreducible representation is
obtained as follows: First we have \be
K_3|k;n_1,n_2\rangle=\frac{1}{2}(n_1+n_2+1)|k;n_1,n_2\rangle
=(k+n)|k;n_1,n_2\rangle \ee and second we have for the Casimir
operator $Q=K_1^2+K_2^2-K_3^2= K_+K_- +K_3(1-K_3)$:
\begin{eqnarray} Q|k;n_1,n_2\rangle &=& \{n_1\,n_2+
\frac{1}{4}(1+n_1+n_2)(1-n_1-n_2)\} |k;n_1,n_2\rangle \nonumber\\
&=&k(1-k)|k;n_1,n_2\rangle~.
\end{eqnarray}
 Thus we have the two relations
\be k=\frac{1}{2}
+\frac{1}{2}|n_1-n_2|~;~~n=\frac{1}{2}(n_1+n_2)-\frac{1}{2}|n_1-n_2|
=\mbox{min}\{n_1,\, n_2 \}\; . \ee They show that in this
construction only representations with half integer or integer
positive $k$ are realizable and that the relations (131) are
symmetric in $n_1$ and $n_2$. The latter property means that,
except for $k=1/2$ where $n_1=n_2$, each irreducible
representation with fixed $k$ occurs twice in the tensor product $
\mbox{$\cal H$}^{osc}_1\otimes \mbox{$\cal H$}^{osc}_2$  (because
$n_1-n_2=\pm(2k-1)$)  of the harmonic oscillator Hilbert spaces
$\mbox{$\cal H$}^{osc}_j, j=1,2$, realized, e.g., by the
orthogonal Hermite functions on $L^2(\Real,dx)$. For $k=1$ we have
the two possibilities $e_{n_1}(x_1)\otimes e_{n_1+1}(x_2)$ or
$e_{n_1}(x_1)\otimes e_{n_1-1}(x_2)$, where $e_{n_1}(x_1) \in
\mbox{$\cal H$}^{osc}_1$ and $e_{n_2}(x_2) \in \mbox{$\cal
H$}^{osc}_2$.
\subsection{Reintroducing Planck's constant} Up to now we have set
$\hbar=1$. We restore it explicitly in the same way as in the case
of the rotation group: We just multiply each operator $K_j$ by
$\hbar$. This corresponds to the fact that the operator
$\hat{p}=K_3$ is canonically conjugate to the dimensionless angle
variable $\varphi$. According to Eq.\ (12) the quantization of the
horizon area  is then given by \be
A_{D-2}(k;n)=(k+n)\,a_{D-2}~,~~a_{D-2}=\frac{32
\pi^2}{\gamma\,(D-3)} \,l_{P,D}^{D-2}. \ee \section{Conclusions}
Our above results show that the original ansatz of Ref.\
\cite{ka1} to associate the Bekenstein spectrum (1) or (2) with a
$finite$
 time interval $\Delta \propto R_S(M)$ which precedes the  collaps of
 the Schwarzschild system to a black hole can be put on more solid grounds:
Implementing the finite time interval by $M$--dependent periodic
  boundary conditions  leads to a phase space  with symplectic
  form $d\varphi\wedge dp$  which is globally diffeomorphic
  to $S^1\times \Real^+$. Such a phase space can be quantized group
   theoretically
  by means of the group $SO^{\uparrow}(1,2)$ (or its covering groups).

  The main advantage of this approach is that it provides a Hilbert
space and the basic self-adjoint operators for quantized
Schwarzschild black holes.

  The crucial point is that the classical variable $p$ is proportional
  to the area $A_{D-2}$ of the black hole horizon in any space-time
  dimension $ D \ge 4$ and that the self-adjoint operator $\hat{p}$ has a
  discrete spectrum in any irreducible unitary representation of
  $SO^{\uparrow}(1,2)$. As we want the spectrum to be positive --- because
 the area $A$ is a positive quantity --- only the positive discrete series
 among the irreducible unitary representations has the required properties.
 It provides the spectrum \be A_{D-2}(k;n) \propto k+n,~n \in
  \Natural_{\, 0}~ .\ee
Here the number $k> 0$ mathematically characterizes the
representation and physically the ``remnant'' area of the ground
state. As the energy of the $n$-th level is given by --- see Eq.\
(10) --- \be E_{k;n} = \alpha_D\, (k+n)^{(D-3)/(D-2)}\;
E_{P,D}~,~n \in \Natural_{\, 0} ~, \ee the number $k$ determines
the ground state energy like the number $1/2$ in the case of the
harmonic oscillator. The value of $k$ depends on the
representation to be employed: For the ``true'' representations of
$SO^{\uparrow}(1,2)$ themselves $k$ can take only the values
$1,2,\ldots$ (corresponding to the $s$-, $p$-, etc.\ states of the
hydrogen atom!). For the two-fold covering groups $SU(1,1)$ or
$SL(2,\Real)$ $k$ may assume the values $k=1/2,1,3/2,2, \ldots$
and for the universal covering group
$\widetilde{SO^{\uparrow}(1,2)}$ $k$ can be any real positive
number. Physical arguments will have to select the right value. If
$k=1$ the ground state area quanta have the same value as those of
the excited levels which would leave us with only one kind of area
quanta, but this must not be so, as can be seen from the harmonic
oscillator where the energy of the ground state is just half of
the basic energy quantum $\hbar \omega$.

In Ref.\ \cite{bo2} arguments will be presented which suggest that
the possible values of $k$ should be restricted to the interval
$(0,1]$.

 As to the physics we see the
following picture emerging: The $area$ of the quantized
Schwarzschild black hole is built up {\em additively} and
equidistantly from basic
 quanta whereas the $energy$ (134) behaves differently: the energy of the
 $n$-th level may be interpreted
 \cite{ka3,ka4,ka5} as the surface energy of a ``bubble'' of $n$  area quanta.

It is an interesting and supporting result that the eigenvalues of
the area operator in the spherically symmetric sector of loop
quantum gravity (without matter) in 3+1 space-time dimensions for
large $n$ are proportional to $n$, too \cite{bo3}.

  As to the degeneracies of the states one sees immediately
   from the formulae above
 that the eigenstates $e_{k,n}$ etc.\ are not degenerate
  in an irreducible representation.

One might think about passing to reducible representations, e.g.
in terms of Fock spaces constructed from ``1-particle'' wave
functions discussed above (2nd quantization). The operator $K_3$
then becomes a sum of the corresponding ``irreducible'' $K_3$. The
degeneracies of the associated eigenvalues are then given by the
possible partitions of a positive number $n$ into smaller ones.
For large $n$ this yields \cite{ra} $d_n \sim g^{ \sqrt{n}}$,
$g>1$, in contrast to $g^n$ required to yield the correct
Bekenstein-Hawking entropy (see Eq.\ (133)).

 Yet one gets the desired thermodynamics, if, as one of us  has
 proposed \cite{ka5},
  each area quantum is
 assigned two degrees of freedom corresponding to the two possible
 orientations of a (classical) sphere.
  The energy spectrum (134) together with the
 degeneracy $2^n$ of the $n$-th level then lead \cite{ka4,ka5}
 to the Hawking temperature
 and the Bekenstein-Hawking entropy of a Schwarzschild black hole!

Thus, altogether a quite coherent picture of the quantum theory of
Schwarz\-schild black holes and their thermodynamics emerges.

The groups $SO^{\uparrow}(1,2),~ SL(2,\Real)$ etc.\ and their Lie
algebra have been playing a number of roles in the context of
recent attempts to quantize black holes:

 Hollmann \cite{hol} has
analyzed the quantum theory of Schwarzschild (Taub-NUT) black
holes in terms of the coset space $SL(2,\Real)/SO(2)$ which yields
a continuous spectrum, whereas we use the coset space
$SL(2,\Real)/\mbox{N}$, where $\mbox{N}$ is the nilpotent group
from an Iwasawa decomposition.

 The group $SL(2,\Real)$ and its Lie algebra have a
prominant role also in recent discussions of black holes in
$(D=2)$- and
 $(D=3)$-dimensional
models of quantum gravity, especially Anti-de Sitter spaces
$AdS_2$ and $AdS_3$ (see Refs. \cite{str,car,ba} and the
literature quoted there). In the 3-dimensional case the Lie
algebra $\mbox{$\cal L$}SL(2,\Real)$ plays an essential role as
the basic subalgebra of the associated Virasoro algebras
\cite{go,ma}. At the moment it is an open question whether and how
these approaches are related to ours above.

See also the interesting application of the group $SL(2,\Real)$ to
black holes by Gibbons and Townsend \cite{gi}.
\section*{Acknowledgement}
We thank N.\ D\"uchting for discussions.

Note added: After our paper was submitted as an e-print we became
aware of an earlier group theoretical quantization of the
symplectic manifold $S^1 \times \mathbb{R}^+$ in terms of the
group $SO^{\uparrow}(1,2)$ by R.\ Loll \cite{lo} in a different
context.
\newpage
\section*{Appendices}
\begin{appendix}
\renewcommand{\theequation}{\thesection.\arabic{equation}}
\setcounter{equation}{0}
 \section{Properties of the group $SO^{\uparrow}(1,2)$, of some
of its covering groups and their irreducible unitary
representations of the positive discrete series}
 The purpose of the present appendix is to summarize
the main properties of the group $SO^{\uparrow}(1,2)$ which are important
for our discussion above, where this group, its covering groups, its Lie
algebra and its irreducible unitary representations, especially those of
the positive discrete series, have been employed as the quantizing framework
for Schwarzschild black holes. Practically all of this appendix is contained
in a wealth of literature about this group which is the most elementary
of noncompact semisimple Lie groups. Potential readers of this paper, however,
will find it convenient to have the required properties assembled in one
unit. \\
The  essential classical paper on the group $SO^{\uparrow}(1,2)$
 and its irreducible unitary
 repesentations is (still!) that of Bargmann \cite{bar1}. In the
 meantime there are a number of monographs which deal with the
 group $SO^{\uparrow}(1,2)$, its covering groups and their representations
 \cite{gel,sa1,wa,lan,hel2,kna,sug,vil,how}. As these textbooks contain many
 references to the original literature we mention only the most essential
 ones for our purposes. \subsection{The group and some of its covering groups}
In order to see the homomorphism between $SO^{\uparrow}(1,2)$ and
its mutually isomorphic twofold covering groups
$SU(1,1),SL(2,\Real)$ and the symplectic group $Sp(1,\Real)$ in 2
dimensions it is convenient to start from the action of the group
$SL(2,\Complex)$
--- the twofold covering group of
 the proper orthochronous
Lorentz group $SL^{\uparrow}(1,3)$ ---  on Minkow\-ski space $M^4$
 with the scalar product
$x\cdot x=(x^0)^2-(x^1)^2-(x^2)^2-(x^3)^2$: Define the hermitean
matrix \be X= \left( \begin{array}{cc} x^0+x^3 & x^1-ix^2 \\
x^1+ix^2& x^0-x^3 \end{array} \right)~,~~ \mbox{det}X =
(x^0)^2-(x^1)^2-(x^2)^2-(x^3)^2~. \ee If $C \in SL(2,\Complex),~
\mbox{det}C=1,$ then \be X \rightarrow \hat{X}=C\cdot X\cdot
C^+~,~~ \mbox{det}\hat{X}= \mbox{det}X~, \ee induces a proper
orthochronous Lorentz transformation on $M^4$. Here $C^+$ means
the hermitean conjugate of the matrix $C$. \\ Subgroups
$SO^{\uparrow}(1,2)$ may be obtained by looking for those
transformations (A.2) which leave one of the coordinates
$x^j,~j=1,2$ or $3$ fixed: \\ The transformations with the
property \be C\cdot \left( \begin{array}{cc} 1 & 0 \\ 0& -1
\end{array} \right)\cdot C^+= \left( \begin{array}{cc} 1 & 0 \\ 0&
-1 \end{array} \right) \ee leave the coordinates $x^3$ invariant
and represent the subgroup $SU(1,1)=\{g_0\} \subset
SL(2,\Complex)$: \be g_0= \left( \begin{array}{cc} \alpha & \beta
\\ \bar{\beta}& \bar{\alpha} \end{array} \right)~,~~
|\alpha|^2-|\beta|^2=1~~. \ee $\bar{\alpha}$: complex conjugate of
$\alpha$. If we let $g_0$ act on a 2-dimensional complex vector
space, then \be g_0 \cdot \left(
\begin{array}{c} z_1 \\ z_2 \end{array} \right)= \left(
\begin{array}{c} \hat{z}_1 \\ \hat{z}_2 \end{array}
\right)~,~~|\hat{z}_1|^2 -|\hat{z}_2|^2 = |z_1|^2-|z_2|^2~. \ee If
now $|z_2|>|z_1|$ and $z=z_1/z_2$ then $SU(1,1)$ maps the interior
$\mbox{$\cal D$}=\{z;|z|<1\}$ of the unit disc in the complex
$z$-plane (transitively) onto itself: \be z \rightarrow
\hat{z}=\frac{\alpha z + \beta}{ \bar{\beta} z +
\bar{\alpha}}~.\ee If we write $\alpha = \alpha_1 +i\alpha_2,
\beta = \beta_1 + i \beta_2$, then $|\alpha|^2-|\beta|^2=
\alpha_1^2+\alpha_2^2-\beta_1^2-\beta_2^2=1. $ This means that the
group manifold of $SU(1,1)$ is homeomorphic to the 3-dimensional
Anti-de Sitter space \cite{ell} $AdS_3$. \\ The subgroup of
$SL(2,\Complex)$ with the property \be C\cdot \left(
\begin{array}{cc} 0 & -i \\ i & 0 \end{array} \right)\cdot C^+=
\left(
\begin{array}{cc} 0 & -i \\ i & 0 \end{array} \right) \ee leaves
the coordinates $x^2$ invariant. It constitutes the group
$SL(2,\Real)$: \be C=g_1= \left( \begin{array}{cc} a_{11} & a_{12}
\\ a_{21} & a_{22} \end{array} \right)~,~ a_{jk}\in
\Real~,~\mbox{det}g_1=1~. \ee As \be g_1\cdot \left(
\begin{array}{cc} 0 & 1 \\ -1 & 0 \end{array} \right)\cdot
g_1^{T}= \left(
\begin{array}{cc} 0 & 1 \\ -1 & 0 \end{array} \right)~, \ee the
group $SL(2,\Real)$ is identical with the real symplectic group
$Sp(1,\Real)$ in 2 dimensions. \\ The unitary matrix \be C_0 =
\frac{1}{\sqrt{2}}\left( \begin{array}{cc} 1 & -i \\ -i & 1
\end{array} \right)~,~\mbox{det}C_0=1~,~C_0^{-1}=
\frac{1}{\sqrt{2}}\left( \begin{array}{cc} 1 & i \\ i & 1
\end{array} \right)=C_0^+~, \ee has the property \be C_0\cdot
\left( \begin{array}{cc} 0 & -i \\ i & 0 \end{array} \right)
C_0^{-1}=  \left( \begin{array}{cc} 1 & 0 \\ 0 & -1 \end{array}
\right) \ee and therefore implements an isomorphism between
$SU(1,1)$ and $SL(2,\Real)$: \be C_0\cdot g_0\cdot C_0^{-1}=g_1~.
\ee It is obvious that the isomorphic groups $SU(1,1),SL(2,\Real)$
and $Sp(1,\Real)$ are twofold covering groups of
$SO^{\uparrow}(1,2)$. \\ The group $SL(2,\Real)$ maps the complex
upper half plane $\Complex^{+i}=\{z=x+iy,~y>0\}$ transi\-tive\-ly
onto itself: \be z \rightarrow \hat{z}=\frac{a_{11}z
+a_{12}}{a_{21}z+a_{22}}~,~~
\Im(\hat{z})=\frac{y}{(a_{22}+a_{21}x)^2+ a_{21}^2y^2}~. \ee Of
special interest for our purposes is the (unique!) Iwasawa
decomposition \cite{hel1,sug}
 of the groups
$\mbox{G}_1\equiv SL(2,\Real)$ and $\mbox{G}_0\equiv SU(1,1)$: $
\mbox{G}_1\equiv
 \mbox{K}_1\cdot \mbox{A}_1\cdot \mbox{N}_1$,
$ \mbox{G}_0=\mbox{K}_0\cdot \mbox{A}_0\cdot \mbox{N}_0$, where K
is the maximal compact subgroup, A a maximally abelian noncompact
subgroup and N a nilpotent group. For $\mbox{G}_1$ this
decomposition is
\begin{eqnarray}\mbox{K}_1: && k_1= \left(
\begin{array}{cc} \cos(\theta/2) & \sin(\theta/2) \\ -\sin(\theta/2) &
\cos(\theta/2) \end{array} \right)~,~ \theta \in (-2\pi,+2\pi]~,
\\ \mbox{A}_1: && a_1 = \left(
\begin{array}{cc} e^{\ds t/2} & 0 \\ 0 &
e^{\ds -t/2} \end{array} \right)~,~t \in \Real~, \\ \mbox{N}_1: &&
n_1 = \left(
\begin{array}{cc} 1 & \xi \\ 0 &
1 \end{array} \right)~,~ \xi \in \Real~~. \end{eqnarray} Each
element $g_1$ has a unique decomposition $g_1=k_1\cdot a_1 \cdot
n_1$. The isomorphism (12) gives the corresponding decomposition
of $\mbox{G}_0$:
 \begin{eqnarray}\mbox{K}_0: && k_0= \left(
\begin{array}{cc}  e^{\ds -i\theta/2} & 0\\  0 &
e^{\ds i\theta/2} \end{array} \right)~,~ \theta \in
(-2\pi,+2\pi]~, \\ \mbox{A}_0: && a_0 = \left(
\begin{array}{cc} \cosh(t/2) & -i\sinh(t/2) \\ i\sinh(t/2) &
\cosh(t/2) \end{array} \right)~,~t \in \Real~, \\ \mbox{N}_0: &&
n_0 = \left(
\begin{array}{cc} 1-i\xi/2 & \xi/2 \\ \xi/2 &
1+i\xi/2 \end{array} \right)~,~ \xi \in \Real~~.
\end{eqnarray} In addition to the above subgroups the following
two ones are of interest to us:
\begin{eqnarray} \mbox{B}_1: && b_1=
\left(
\begin{array}{cc} \cosh(s/2) & \sinh(s/2) \\ \sinh(s/2) &
\cosh(s/2) \end{array} \right)~,~s \in \Real~, \\ \mbox{B}_0: &&
b_0 =C_0^{-1}\cdot b_1 \cdot C_0 = \left(
\begin{array}{cc} \cosh(s/2) & \sinh(s/2) \\ \sinh(s/2) &
\cosh(s/2)
 \end{array} \right)~, \\ \bar{\mbox{N}}_1: && \bar{n}_1 =
\left(
\begin{array}{cc} 1 & 0 \\ \xi &
1 \end{array} \right)~,~ \xi \in \Real~, \\ \bar{\mbox{N}}_0: &&
\bar{n}_0= \left(
\begin{array}{cc} 1+i\xi/2 & \xi/2 \\ \xi/2 &
1-i\xi/2 \end{array} \right)~. \end{eqnarray} Two more
decompositions of $SL(2,\Real)$ or $SU(1,1)$ are important for the
construction of their unitary representations:
\\ Cartan (or ``polar'') decomposition \cite{hel1,sug}: \\ Each
element of $SL(2,\Real)$ can be written as \be g_1=  k(\theta_2)
\cdot a_1(t)\cdot k(\theta_1)~, \ee where $a_1(t)$ is determined
uniquely and $k(\theta_1),k(\theta_2)$ up to a relative sign, that
is up to the center $\Integer_2$ of $SL(2,\Real)$. \\ Bruhat
decomposition \cite{hel1,wa,lan}: \\ From \begin{eqnarray}
\lefteqn{ k(\theta)\cdot a_1(t)\cdot k(-\theta)=} \hspace{-0.4cm}
 \\ &&  \left( \begin{array}{cc}
\cos^2(\theta/2)e^{t/2}+\sin^2(\theta/2)e^{-t/2} &
\sin(\theta/2)\cos(\theta/2)(e^{-t/2}-e^{t/2}) \\
\sin(\theta/2)\cos(\theta/2)(e^{-t/2}-e^{t/2}) &
\cos^2(\theta/2)e^{-t/2}+\sin^2(\theta/2)e^{t/2} \end{array} \right)
\nonumber \end{eqnarray}
one sees that \begin{eqnarray}
 k(\theta)\cdot a_1(t)\cdot k(-\theta)&=& a_1(t)~ \mbox{for}~
\theta=0, 2\pi~, \\ k(\theta)\cdot a_1(t)\cdot k(-\theta)& \subset
& \mbox{A}_1 ~\mbox{for}~ \theta=0,\pm \pi, 2\pi~, \end{eqnarray}
 which means that the centralizer
$C_{\mbox{K}_1}(\mbox{A}_1)$ and normalizer
$N_{\mbox{K}_1}(\mbox{A}_1)$ of $\mbox{A}_1$ in $\mbox{K}_1$ are
given by \be C_{\mbox{K}_1}(\mbox{A}_1)=\left\{ \pm \left(
\begin{array}{cc} 1 & 0 \\ 0 & 1 \end{array} \right)
\right\}=\Integer_2~,~~ \ee \be N_{\mbox{K}_1}(\mbox{A}_1) =
\left\{ \pm \left(
\begin{array}{cc} 1 & 0 \\ 0 & 1
\end{array} \right),~
 \pm \left( \begin{array}{cc}0  & 1 \\ -1 & 0 \end{array}
\right) \right\}~. \ee
 The quotient group \begin{eqnarray} W&=&N_{K_1}(\mbox{A}_1)/\Integer_2\\&& =
\left\{  \left( \begin{array}{cc} 1 & 0 \\ 0 & 1 \end{array}
\right)\mbox{ mod }\Integer_2,~w=
 \left( \begin{array}{cc} 0  & 1 \\ -1 & 0 \end{array}
\right)\mbox{ mod }\Integer_2
  \right\} \nonumber \end{eqnarray}
   is called the Weyl-group of $SL(2,\Real)$. Its associated Bruhat
  decomposition of $SL(2,\Real)$ is
  \be \mbox{G}_1=\Integer_2\cdot \mbox{A}_1 \cdot \mbox{N}_1 \cup
   \mbox{N}_1\cdot
   w \cdot \Integer_2 \cdot \mbox{A}_1
   \cdot \mbox{N}_1~,~ w=
 \left( \begin{array}{cc} 0  & 1 \\ -1 & 0 \end{array}
\right)~.
  \ee  Here $\Integer_2\cdot \mbox{A}_1$ is the group \be
  \mbox{D}_1=\Integer_2\cdot
  \mbox{A}_1 = \left\{ \left(
  \begin{array}{cc} c &0 \\ 0 & c^{-1} \end{array} \right)~,~
   c \in \Real -\{0\}
  \right\}~. \ee The relation (A.31) means that each element of $SL(2,\Real)$
  may be decomposed uniquely either as an element of the ``parabolic''
  subgroup $\mbox{P}_1 =\mbox{D}_1\cdot \mbox{N}_1$
   or as
  an element of $\mbox{N}_1\cdot w \cdot \mbox{P}_1$. \\
The Bruhat decomposition of $SL(2,\Real)$ plays a central role in
Sally's construction \cite{sa1} of the irreducible unitary
representations of the universal covering group
$\widetilde{SL(2,\Real)}$. \\ As the compact subgroup
$\mbox{K}_1$,
 or $\mbox{K}_0,\; \simeq S^1$, is infinitely connected,
  the groups $SL(2,\Real)$ and
  $SU(1,1)$
 have an infinitely sheeted universal covering group which, according to
 Bargmann, may be parametrized as follows: Starting from $SU(1,1)$ define
 \begin{eqnarray} \gamma& =& \beta/\alpha~,~|\alpha|^2-|\beta|^2=1~(
 \Rightarrow |\gamma|<1);~
  ~\omega = \arg(\alpha)~;
 \\ \alpha & =& e^{\ds i\omega}(1-|\gamma|^2)^{-1/2}~,~ |\gamma|<1~,~\beta =
e^{\ds i\omega}\gamma(1-|\gamma|^2)^{-1/2}~. \end{eqnarray} Then
\begin{eqnarray} SU(1,1)& =& \{g_0=(\omega,\gamma),~\omega \in
 (-\pi, \pi],~|\gamma|<1 \}~,~ \\ \tilde{\mbox{G}}\equiv \widetilde{SU(1,1)}
 &=&
 \widetilde{SL(2,\Real)}=  \{\tilde{g}=(\omega,\gamma),~\omega \in \Real,~
 |\gamma|<1 \}~. \end{eqnarray} The group composition law for $\tilde{g}_3 =
 \tilde{g}_2 \cdot \tilde{g}_1$ is given by \begin{eqnarray} \gamma_3 & = &
 (\gamma_1 +\gamma_2 e^{\ds
 -2i\omega_1})(1+\bar{\gamma}_1\gamma_2e^{\ds -2i\omega_1})^{-1}~~, \\
 \omega_3&=& \omega_1 + \omega_2 +\frac{1}{2i}\ln[(
1+\bar{\gamma}_1\gamma_2e^{\ds - 2i\omega_1})(
1+\gamma_1\bar{\gamma}_2e^{\ds 2i\omega_1})^{-1}]~. \end{eqnarray}
\subsection{Lie algebra} As the structure of the 3-dimensional
Lie algebra $\mbox{$\cal L$} SO^{\uparrow}(1,2)= \{l\}$ of
$SO^{\uparrow}(1,2)$ is the same as that of all its covering
groups we may calculate it by using any of them. For $SL(2,\Real)$
we get from the subgroups (A.14)-(A.16), (A.20) and (A.22): \be
l_{K_1}= \frac{1}{2}\left( \begin{array}{cc} 0& 1\\-1&0
\end{array} \right),~ l_{A_1}= \frac{1}{2}\left(
\begin{array}{cc} 1& 0\\0&-1 \end{array} \right),~ l_{B_1}=
\frac{1}{2}\left(
\begin{array}{cc} 0& 1\\1&0 \end{array} \right)~, \ee \be l_{N_1}=
\left( \begin{array}{cc}0 & 1\\0&0 \end{array} \right)~,~~
l_{\bar{N}_1}= \left( \begin{array}{cc}0 & 0\\1&0
\end{array} \right)~, \ee which are not independent
(in the following we skip the indices ``1'' or ``2'', because the
structure relations are independent of them): \be l_{N} +
l_{\bar{N}}= 2\; l_{B}~,~~~ l_{N} - l_{\bar{N}}= 2\; l_{K}~~. \ee
We have the commutation relations
\begin{eqnarray} [l_{K},l_{A}]&=&-l_{B}~,~~[l_{K},l_{B}]=l_{A}~,
~~[l_{A},l_{B}]
=l_{K}~, \\ &&
[l_{K},l_{N}]=l_{A}~,~~[l_{K},l_{\bar{N}}]=l_{A}~,\\
&&[l_{A},l_{N}]=l_{N}~,~~ [l_{A},l_{\bar{N}}]=-l_{\bar{N}}~,
\\ &&[l_{B},l_{N}]=-l_{A}~,~~
[l_{B},l_{\bar{N}}]=l_{A}~,\\ && [l_{N},l_{\bar{N}}]=2\; l_{A}~.
\end{eqnarray} The relations (A.42) show that the algebra is
semisimple, the first of the Eqs.\ (A.44) that $\mbox{A}$ and
$\mbox{N}$ combined form a 2-dimensional subgroup and that
$\mbox{A}$ is a normalizer of $\mbox{N}$.
\subsection{Irreducible unitary representations \\
of the positive discrete series} As the group $SO^{\uparrow}(1,2)$
is noncompact, its irreducible unitary representations are
infinite-dimensional. Their structure can be seen already from its
Lie algebra: In unitary representations the elements
$-il_{K},-il_{A},-il_{B}$ of the Lie algebra correspond to
self-adjoint operators $K_3,K_1,K_2$ which obey die commutation
relations \be
[K_3,K_1]=iK_2~,~~[K_3,K_2]=-iK_1~,~~[K_1,K_2]=-iK_3~, \ee or,
with the definitions \be K_+=K_1+iK_2~,~~K_-=K_1-iK_2~, \ee \be
[K_3,K_+]=K_+~,~~[K_3,K_-]=-K_-~,~~[K_+,K_-]=-2K_3~. \ee The
relations (A.47) are invariant under the replacement $K_1
\rightarrow -K_1, K_2 \rightarrow -K_2$ and the relations (A.49)
invariant under $K_+ \rightarrow \omega K_+, K_- \rightarrow
\bar{\omega} K_-,$ where $ |\omega|=1$. These relations are in
addition invariant under the transformations $ K_+ \leftrightarrow
K_-, K_3 \rightarrow -K_3$.

 In irreducible unitary representations the
operator $K_-$ is the adjoint operator of $K_+:\;
(f_1,K_+f_2)=(K_-f_1,f_2)$,
 and vice versa, where it is assumed that $f_1,f_2$ belong to the domains
 of definition of $K_+$ and $K_-$.

The Casimir operator $Q$ of a representation is defined by \be
Q=K_1^2+K_2^2-K_3^2 \ee and we have the relations \be
K_+K_-=Q+K_3(K_3-1)~,~~K_-K_+=Q+K_3(K_3+1)~. \ee All unitary
representations make use of the fact that $K_3$ is the generator
of a compact group and that its eigenfunctions $g_m$ are
normalizable elements of the associated Hilbert space $\cal H$.

The relations (A.49) imply \begin{eqnarray} K_3\,g_m&=&m\,g_m~, \\
K_3K_+g_m&=&(m+1)K_+g_m~, \\ K_3K_-g_m &=& (m-1)K_-g_m~,
\end{eqnarray} which, combined with Eqs.\ (A.51), lead to
\begin{eqnarray} (g_m,K_+K_-g_m)&=&(K_-g_m,K_-g_m)=q+m(m-1) \ge
0~, \\ (g_m, K_-K_+g_m)&=&q+m(m+1)\ge 0 ~, ~~q=(g_m,Qg_m),
\end{eqnarray} implying \be (K_+g_m,K_+g_m)= 2m+(K_-g_m,K_-g_m)
\ge 0. \ee In the following we assume that we have an irreducible
representation for which the functions $g_m$ are eigenfunctions of
the Casimir operator $Q$, too: $Qg_m=q\,g_m$.

 The relations
(A.52)-(A.54) show that the eigenvalues of $K_3$ in principle can
be any real number, where, however, different eigenvalues
 differ by an integer. For the ``principle'' and the ``complementary'' series
the spectrum of $K_3$ is unbounded from below and above .

 The
positive discrete series $D_+$ of irreducible  unitary
representations is
 characterized by the property that there exists a lowest eigenvalue
 $m=k$ such that \be K_-g_k=0~. \ee Then the relations (A.55)-(A.57) imply
 \be q=k(1-k)~,~~~ k>0~,~~m=k+n,~ n=0,1,2,\ldots . \ee The relations
 (A.52)-(A.54) now take the form \begin{eqnarray} K_3g_{k,n}&=&(k+n)g_{k,n}~,
 \\
 K_+g_{k,n}&=&\omega_n\,[(2k+n)(n+1)]^{1/2}g_{k,n+1}~,~~|\omega_n|=1~, \\
 K_-g_{k,n}&=&\frac{1}{\omega_{n-1}}[(2k+n-1)n]^{1/2}g_{k,n-1}~.\end{eqnarray}
 The phases $\omega_n$ guarantee that $(f_1,K_+f_2)=(K_-f_1,f_2)$. \\
 Up to now we have allowed for any value of $k>0$. It turns out
 \cite{puk,sa1} that this
 is so for the irreducible representations of the universal covering
 group $\widetilde{SO^{\uparrow}(1,2)}$. These representations may
 be realized for $k \ge 1/2$ in the Hilbert space of holomorphic
  functions on the unit
 disc $\mbox{$\cal D$}=\{z,|z|<1\}$ with the scalar product \be
 (f,g)_k=\frac{2k-1}{\pi}\int_{\mbox{$\cal D$}}\bar{f}(z)g(z)
 (1-|z|^2)^{2k-2}dxdy~. \ee
 as \begin{eqnarray}
 [U(\tilde{g},k)f](z)&=&e^{\ds 2ik\omega}(1-|\gamma|^2)^{\ds k}
 (1+\bar{\gamma}z)^{\ds -2k}
 f\left( \frac{\alpha z+\beta}{\bar{\beta}z+\bar{\alpha}}\right)\; , \\
  \tilde{g}&=&(\omega,\gamma)~,~~ \left( \begin{array}{cc} \alpha & \beta \\
 \bar{\beta}& \bar{\alpha} \end{array} \right)~=~h(\tilde{g})~
  \in SU(1,1)~. \end{eqnarray}
Because $|\gamma\,z|<1$, the function $(1+\bar{\gamma}z)^{-2k}$ is, for $k>0$,
 defined
in terms of a series expansion.

 For $SU(1,1)$ we have $\omega \in
\Real \bmod 2\pi$. Uniqueness of the phase factor then requires
$k=1/2,1,3/2,\cdots$.

 For $SO^{\uparrow}(1,2)$ itself we
have $\omega \in \Real \bmod \pi$ which implies $k=1,2,\cdots$.

 More about the unitary representations can be
found in chapter 5 of the main text.
\section{Reduced phase space  of $D$-dimensional \\ Schwarzschild
gravitational systems}
\setcounter{equation}{0}
\subsection{Symplectic reduction}
We start from the spherically symmetric {\sc ADM} line-element
\begin{equation}\label{ds}
  ds^2=-N^2dt^2+P^2(dr+N_rdt)^2+Q^2d\Omega^2~,
\end{equation}
where $r$ is the radial coordinate and $d\Omega^2$ the line
element of a unit sphere $S^{D-2}$, embedded in
$(D-1)$-dimensional flat space. The functions $P$, $Q$, $N$ and
$N_r$ depend only on $r$ and $t$.

 The line element of the
two-dimensional radial manifold with coordinates $(x^0,x^1)=(t,r)$
will be denoted by
\begin{equation}
  d\sigma^2=g_{ij}dx^idx^j=-N^2dt^2+P^2(dr+N_rdt)^2.
\end{equation}
The line element $ds^2$ is conformally equivalent to
\begin{eqnarray} d\hat{s}^2&=&d\hat{\sigma}^2+d\Omega^2~,~~~
 d\hat{\sigma}^2=Q^{-2}d\sigma^2~, \\ && d\Omega^2=g_{AB}\;dy^Ady^B, ~
 2\leq A,B\leq D-1 ~, \nonumber \end{eqnarray}
where $d\Omega^2$ is the line element
 of a
$(D-2)$-dimensional space of constant positive curvature $1$. Its
Ricci curvature tensor is given by $R^{(D-2)}_{AB}=\Lambda
g_{AB}$. A direct calculation using, e.~g., the special form $$
d\Omega^2=\frac{\delta_{CD}\,y^Cy^D}{(1+\frac{1}{4}\delta_{AB}\,y^Ay^B)^2}
$$ shows that $\Lambda=D-3$.

 In order to perform a symplectic
reduction we have to insert the metric (\ref{ds}) into the
Einstein-Hilbert-action. This has already been done, for a
different purpose, in Ref.\ \cite{sch} with the result that the
spherically symmetric theory is described by a two-dimensional
dilaton action
\begin{equation}
  S=\int dtdr\sqrt{|\det \tilde{g}_{ij}|}\left(\phi \tilde{R}-V(\phi)\right).
\end{equation}
The metric $\tilde{g}_{ij}$ is obtained from $g_{ij}$ by the conformal
transformation \be
d\tilde{\sigma}^2=\tilde{g}_{ij}dx^idx^j=Q^{D-3}d\sigma^2~,\ee and the
dilaton field $\phi$ is introduced by \be \phi=Q^{D-2}~,\ee with potential
\be V(\phi)=-(D-2)\Lambda\phi^{-(D-2)^{-1}}=-(D-2)(D-3)\phi^{-(D-2)^{-1}}~.
\ee
Arrived at a two-dimensional dilaton action we can adopt  results
of the symplectic reduction of Ref.\ \cite{kun}. There the line element
is parameterized as
\begin{equation}
  d\tilde{\sigma}^2=e^{2\rho}(-\sigma^2dt^2+(dr+L\,dt)^2)
\end{equation}
and the canonical variables are
$(\rho,P_{\rho};\phi,P_{\phi};L,P_L;\sigma,P_{\sigma})$ subject to the
constraints
\begin{eqnarray} P_L &\approx & 0~, \\ P_{\sigma}&\approx &0 ~, \\
  {\cal E}:=\rho'P_{\rho}+\phi'P_{\phi}-P_{\rho}' &\approx &0~,\\
  {\cal G}:=2\phi''-2\phi'\rho'-\frac{1}{2}P_{\phi}P_{\rho}+e^{2\rho}V(\phi)
        & \approx& 0~.
\end{eqnarray}
A prime denotes differentiation with respect to $r$, whereas a dot
will denote differentiation with respect to $t$.

 The first two
constraints eliminate the variables $L$ and $\sigma$, whereas the
remaining two are equivalent to ${\cal E}\approx0$ and
$C'\approx0$ with
\begin{equation}\label{C}
  C:=e^{-2\rho}\left(\frac{1}{4}P_{\rho}^2-(\phi')^2\right)-j(\phi)~,~~
  \frac{dj}{d\phi}=V(\phi)~.
\end{equation}
$C$ is a physical observable, which is forced by the constraints to be
spatially constant.  It represents the only physical degree of freedom. \\
The last step consists in connecting the results of the two quoted papers
\cite{kun,sch} by the
transformation of variables
\begin{eqnarray}
  \rho &\mapsto & P=e^{\ds \rho}\phi^{\ds (3-D)/(2D-4)}~, \\
  \phi &\mapsto& Q  =   \phi^{\ds 1/ (D-2)}~, \\
  L    &\mapsto & N_r= L~, \\
  \sigma &\mapsto & N= ~ \sigma e^{\ds \rho}\phi^{\ds(3-D)/(2D-4)}~.
\end{eqnarray}
For the  momenta we get
\begin{eqnarray}  P_N&=&0~,\\ P_{N_r}&=&0~, \\
  P_{\phi} & = &\frac{1}{D-2}Q^{\ds 3-D}P_Q - \frac{1}{2}\frac{D-3}{D-2}
   PQ^{\ds 2-D}P_P~,\\
  P_{\rho} & = &PP_P=2(D-2) PQ^{\ds D-3}N^{-1}(-\dot{Q}+N_rQ').
\end{eqnarray}
Using our $V(\phi)$ we have \be j(\phi)=-(D-2)^2\phi^{\ds
(D-3)/(D-2)}+k~.\ee Here $k$ is a constant of integration.
Inserting all these formulae in Eq.\ (\ref{C}) we obtain
\begin{equation}
  C=\frac{1}{4}Q^{\ds D-3}P_P^2-(D-2)^2Q^{\ds D-3}\left((Q'P^{-1})^2
  -1\right)-k.
\end{equation}
In order to reveal its relation to the Schwarzschild  mass $M$ we
compare it with the Schwarzschild line element in $D$ space-time
dimensions \cite{per1}:
 $Q=r$, $N_r=0$, and
\begin{equation}
  P^{-2}=N^2=F:=1-\frac{16\pi\, G\,M}{(D-2) \omega_{D-2}\,
  r^{D-3} }~,
\end{equation}
where $G_D$ is the Newton constant, $M$ the Schwarzschild (ADM)
mass.

 Due to $\dot{Q}=N_r=0$ we have $P_P=0$ (outside the
horizon) and the resulting expression
\begin{equation}\label{mass}
  C=-(D-2)^2\,r^{D-3}\,(F-1)-k=\frac{16\pi G_D\,M\,(D-2)}{\omega_{D-2}}-k
\end{equation}
 shows the relation of $C$ to the Schwarzschild mass.
\subsection{Reduced Hamiltonian}
In order to determine the Hamiltonian associated with
 the observable $C$ we have to add an appropriate
surface term which
renders the action functionally differentiable. For our functions
$(\rho,P_{\rho},\phi,P_{\phi},L,\sigma)$ we choose fall-off conditions which
are prescribed by asymptotic flatness of the
Schwarzschild space-time. The remaining formulae remain, however, true for
arbitrary dilaton potentials $V(\phi)$.
For the functions $(P,P_P,Q,P_Q,N,N_r)$ the fall-off conditions at
$r\to\pm\infty$ ($r$ is the radial coordinate of an asymptotically cartesian
coordinate system) are
\begin{eqnarray}
  Q &=& |r|+O(|r|^{-\epsilon}),\\
  P &=& 1+M_{\pm}(t)|r|^{3-D}+O(|r|^{3-D-\epsilon}),\\
  P_Q &=& O(|r|^{-1-\epsilon}),\\
  P_P &=& O(|r|^{-\epsilon}),\\
  N_r &=& O(|r|^{-\epsilon}),\\
  N &=& N_{\pm}(t)+O(|r|^{-\epsilon}).
\end{eqnarray} From
 these conditions follow the fall-off conditions for the
functions $\rho$, $P_{\rho}$, $\phi$, $P_{\phi}$, $L$, $\sigma$ by
using the transformation (B.14)-(B.17).

 In order to extract the
surface term we first perform, following \cite{ku}, an appropriate
canonical transformation which replaces the variable $\rho$ by the
observable $C$. The momentum conjugate to $C$ is (locally)
\begin{equation}
  P_C=-\frac{2e^{2\rho}P_{\rho}}{P_{\rho}^2-4(\phi')^2}.
\end{equation} The relation (B.32) becomes singular on the
horizon. One can avoid this by defining a variable conjugate to
$C$ which is nowhere singular, e.g. in the framework of Poisson
$\sigma$--models \cite{psm,st2}, that is the singularity in Eq.\
(B.32) causes no problems here.

 Because $C$ and $P_C$ do not
depend on $P_{\phi}$ we have $\{C,\phi\}=0=\{P_C,\phi\}$ and we
can use $\phi$ as a second variable. However, due to
$\{P_{\phi},C\}\not=0$, we have to replace $P_{\phi}$ by a new
momentum $P_{\psi}$ conjugate to $\psi:=\phi$. As in Ref.\
\cite{ku} it can be calculated by equating the generators of
diffeomorphisms in both sets of variables:
\begin{equation}
  \rho'P_{\rho}+\phi'P_{\phi}-P_{\rho}'=C'P_C+\phi'P_{\psi}
\end{equation}
using the fact that both $C$ and $\phi$ are scalars. The last equation yields
\begin{equation}
  P_{\psi}=P_{\phi}+4\frac{\phi'P_{\rho}'-\phi''P_{\rho}}{P_{\rho}^2-
     4(\phi')^2}
           -2\frac{e^{2\rho}P_{\rho}V(\phi)}{P_{\rho}^2-4(\phi')^2}
          =-2(P_{\rho}^2-4(\phi')^2)^{-1}(P_{\rho}{\cal G}+2\phi'{\cal E})
\end{equation}
which shows that $P_{\psi}$ is constrained to be zero.

 Now we
will show that the transformation
$(\rho,P_{\rho},\phi,P_{\phi})\mapsto(C,P_C,\phi,P_{\psi})$ is
canonical by showing that the difference of the corresponding
Liouville forms is exact. First, we have
\begin{eqnarray}
  && P_{\rho}\delta\rho+P_{\phi}\delta\phi-P_C\delta
  C-P_{\psi}\delta\phi~~~~
   \\
   && ~~~=\delta\left(P_{\rho}+2\phi'\log\left|\frac{\ds 2\phi'
   -P_{\rho}}{\ds 2\phi'+
    P_{\rho}}\right|\right)
    +2\left(\delta\phi\log\left|\frac{\ds 2\phi'+P_{\rho}}{\ds 2\phi'-
    P_{\rho}}\right|\right)'.  \nonumber
\end{eqnarray}
As a consequence of the fall-off conditions the integral over the last
 derivative term
vanishes (at the horizon, i.~e., at the singular points of the
logarithm, the integral has to be interpreted as principal value).
Therefore, the difference
\begin{equation}
  \int\! dr(P_{\rho}\delta\rho+P_{\phi}\delta\phi-P_C\delta C -
    P_{\psi}\delta\phi)
    =\delta\!\!\int\! dr\!\left(\!P_{\rho}+2\phi'\log\left|\frac{2\phi'-
    P_{\rho}}{2\phi'+P_{\rho}}\right|\right)
\end{equation}
of the Liouville forms is exact, and the transformation to the variables
$(C,P_C,\phi,P_{\psi})$ is canonical.
Assuming $\phi'\not=0$, which will be fulfilled for Schwarzschild systems,
the constraints
${\cal E}\approx0$ and ${\cal G}\approx0$ are equivalent to $C'\approx0$ and
$P_{\psi}\approx0$.
We use these constraints together with the new canonical variables and obtain
the action
\begin{equation}
  S=\int dtdr(P_C\dot{C}+P_{\psi}\dot{\phi}-N^CC'-N^{\psi}P_{\psi})~,
\end{equation}
where $N^C$ and $N^{\psi}$ are new Lagrange multipliers. Because of
 the asymptotic
relation
$C'\sim\mp(D-2){\cal G}$, which follows from the fall-off conditions, the
asymptotic values
of $N^C$ and $N$ are related by $N^C\sim\mp(D-2)^{-1}N$.
The only spatial derivative in the new action appears in the
 term $-N^CC'$, and,
therefore,
the action can be made functionally differentiable by adding the surface term
\begin{equation}
  \int dt(N^C_+C_+-N^C_-C_-)=-\int dt(D-2)^{-1}(N_+C_++N_-C_-).
\end{equation}
$C_{\pm}(t)$, $N^C_{\pm}(t)$ and $N_{\pm}(t)$ are the values of
$C$, $N^C$ and $N$ at $r\to\pm\infty$. This surface action and the
formula (\ref{mass}) show that in any space-time dimension the
reduced Hamiltonian of a spherically symmetric black hole is given
by its mass (times the lapse function) after imposing the
constraints: Rescaling $S$ by $(16\pi G_D)^{-1}\omega_{D-2}$ to
obtain the physical action in $D$ dimensions, and setting $k=0$ in
Eq.\ (\ref{mass}), we finally obtain
\begin{equation}
  H_{\mbox{red}}=\frac{\omega_{D-2}}{16\pi\, G_D}(N_++N_-)
  \frac{C}{D-2}=(N_++N_-)\,M
\end{equation}
with $C:=C_+=C_-$ (due to $C'=0$).

\end{appendix}

\end{document}